\documentclass[]{pasj01}
\draft

\Received{}%{yyyy/mm/dd}
\Accepted{}%{yyyy/mm/dd}
\begin{document} 

\title{ 
%\LETTERLABEL %%% <-- uncomment for LETTER article  
%\REVIEWLABEL %%% <-- uncomment for REVIEW article  
Spectral Break of Energetic Pulsar Wind Nebulae Detected with Wideband X-ray Observations}
%Wideband X-ray Spectroscopy of Energetic Pulsar Wind Nebulae}

%%% begin:list of authors
% Do NOT capitalize all letters in "textsc".
\author{Aya \textsc{Bamba}\altaffilmark{1,2}}
\author{Shinpei \textsc{Shibata}\altaffilmark{3}}
\author{Shuta J. \textsc{Tanaka}\altaffilmark{4,5}}
\author{Koji \textsc{Mori}\altaffilmark{6}}
\author{Hiroyuki \textsc{Uchida}\altaffilmark{7}}
\author{Yukikatsu \textsc{Terada}\altaffilmark{8}}
\author{Wataru \textsc{Ishizaki}\altaffilmark{9}}
\altaffiltext{1}{Department of Physics, Graduate School of Science,
the University of Tokyo, 7-3-1 Hongo, Bunkyo-ku, Tokyo 113-0033, Japan}
\email{bamba@phys.s.u-tokyo.ac.jp}
\altaffiltext{2}{Research Center for the Early Universe, School of Science, The University of Tokyo, 7-3-1
Hongo, Bunkyo-ku, Tokyo 113-0033, Japan}
\altaffiltext{3}{School of Science and Technology, Yamagata University, 1-4-12 Kojirakawa-machi, Yamagata, Yamagata 990-8560, Japan}
\altaffiltext{4}{Department of Physical Sciences, Aoyama Gakuin University, 5-10-1 Fuchinobe, Sagamihara, Kanagawa 252-5258, Japan}
\altaffiltext{5}{Graduate School of Engineering, Osaka University, 2-1 Yamadaoka, Suita, Osaka, 565-0871, Japan}
\altaffiltext{6}{Department of Applied Physics and Electronic Engineering, Faculty of Engineering, University of Miyazaki, 1-1 Gakuen Kibanadai-Nishi, Miyazaki 889-2192, Japan}
\altaffiltext{7}{Department of Physics, Kyoto University, Kitashirakawa Oiwake-cho, Sakyo, Kyoto 606-8502, Japan}
\altaffiltext{8}{Graduate School of Science and Engineering, Saitama University, 255 Shimo-Ohkubo, Sakura, Saitama 338-8570, Japan}
\altaffiltext{9}{Center for Gravitational Physics, Yukawa Institute for Theoretical Physics, Kyoto University, Kitashirakawa-Oiwake-Cho, Sakyo-ku, Kyoto, 606-8502, Japan}

%%% end:list of authors

%% `\KeyWords{}' always has to be placed before ``\maketitle'' 
%%  List of Key Words:  https://academic.oup.com/pasj/pages/Pasj_Keywords 
\KeyWords{%
ISM: individual objects (N157B, PSR~J1813$-$1749, PSR~J1400$-$6325, G21.5$-$0.9, Crab Nebula) ---
stars: neutron ---
X-rays: ISM ---
radiation mechanisms: general}

\maketitle

\begin{abstract}
Pulsar wind nebulae (PWNe) are one of the most energetic galactic sources
with bright emissions from radio waves to very high-energy gamma-rays.
We perform wideband X-ray spectroscopy of four energetic PWNe,
N157B, PSR~J1813$-$1749, PSR~J1400$-$6325, and G21.5$-$0.9,
with the Suzaku, Chandra, NuSTAR, and Hitomi observatories.
A significant spectral break or cutoff feature is found in the hard X-ray band for all the samples, except for N157B.
The break energies in the broken power-law fitting are in the range of 4--14~keV, whereas the cutoff energies in the cutoff power-law fitting are at 22~keV or higher.
The break or cutoff energy does not show a significant correlation with either the spin-down energy or characteristic age of the hosting pulsars. 
A possible correlation is found between the photon index change in
the broken power-law fitting and the X-ray emitting efficiency of the pulsars,
although its significance is not high enough to be conclusive.
We discuss what determines the break parameters based on simple models.
\end{abstract}

\section{Introduction}

Relativistic pulsar winds ejected by active pulsars 
form pulsar wind nebulae (PWNe).
High-energy electrons and positrons in the pulsar winds 
are terminated by a strong shock \citep{pacini1973,rees1974,kennel1984}
at 0.01--0.1~pc from a central pulsar \citep{ng2008,bamba2010a},
diffuse out in the downstream of the shock,
and emit electromagnetic waves from radio waves to gamma-rays via synchrotron or inverse Compton processes.
In addition to producing these bright nonthermal emissions, PWNe are one of the most energetic particle accelerators in our galaxy.
Thus, deriving their physical parameters from the wideband spectra is important for understanding acceleration parameters, such as the energy spectra of the accelerated particles and the magnetic field strength.
The current status of the theoretical models and the observations are reviewed in several articles \citep[for example]{gaensler2006,kargaltsev2015,slane2017,amato2020}.

One of the least understood features in the wideband spectra of PWNe is a possible break in the synchrotron emissions in the soft to hard X-ray band.
\citet{tsujimoto2011} performed a cross calibration of X-ray satellites with the bright PWN G21.5$-$0.9 and found that its photon index above 10~keV is significantly larger than that below 10~keV.
This result was confirmed by the NuSTAR observatory \citep{nynka2014} and later the Hitomi \citep{hitomi2018} observatory.
Several PWNe also show a similar spectral break
(Crab: \cite{madsen2015}; PSR~J1400$-$6325: \cite{renaud2010}).
These spectral break features are expected to contain important information because they are often formed by the synchrotron cooling process
\citep{reynolds1984}.
However, the origin of the break is remains unclear;
the break at $\sim 10$~keV cannot be reproduced by one-zone adiabatic models \citep{tanaka2011},
which require a very low magnetic field of $\sim 10~\mu$G \citep{nynka2014}.
Even if we consider multi-zone components,
PWNe often show softening on the outer part of the nebulae \citep{mori2004,guest2019,hu2022},
and thus the spectral energy distribution, including the break, cannot be reproduced well \citep{hitomi2018}.
\citet{hattori2020} applied a time-dependent broadband spectral model of the PWN developed by \citet{gelfand2009} to G21.5-0.9.
They showed that the spectral break feature in the X-ray band can be fitted with a spectral cutoff corresponding to the maximum electron energy.
However, the number of samples is still limited,
and we need more samples to discuss the origin of the spectral break in the X-ray band.

To determine the break energy precisely,
we need the wideband capability of X-ray spectroscopy.
Therefore, we perform wideband spectroscopy of several young and energetic PWNe.
\S\ref{sec:selection} describes the selection of targets and the X-ray dataset as well as the data reduction.
The analysis results for each target are described in \S\ref{sec:result} and they are compared in \S\ref{sec:comparison}.
We discuss our results in \S\ref{sec:discuss}.

\section{Target Selection, Observations, and Data Reduction}
\label{sec:selection}

We investigate the X-ray spectra of PWNe in both the soft and hard X-ray bands.
Our primary observatory in the hard X-ray band is NuSTAR,
which has sufficient spatial resolution to resolve nearby sources in this band \citep{harrison2013}.
NuSTAR has lower sensitivity in the soft X-ray band;
thus, we need data in the soft X-ray band
with a similar point spread function and larger effective area compared with NuSTAR.
Consequently, our primary detector for soft X-rays is X-ray Imaging Spectrometer (XIS) \citep{koyama2007} onboard Suzaku \citep{mitsuda2007}.
For the target that is not observed with Suzaku (PSR~J1400$-$6325),
we apply Chandra data.
We also add a deep observation with Hitomi for one object (G21.5$-$0.9), because Hitomi covers both soft and hard X-ray energy bands
\citep{takahashi2016}, instead of using NuSTAR and Suzaku.

We select four energetic pulsars to investigate the wideband spectroscopy
from the Australia Telescope National Facility (ATNF) Pulsar Catalog,
which contains archival data in both the soft and hard X-ray bands,
as shown in Table~\ref{tab:properties}.
Among the top seven energetic pulsars in the catalog, we cover four objects,
N157B, PSR~J1813$-$1749, PSR~J1400$-$6325, and G21.5$-$0.9, which have the first, fourth, fifth, and seventh largest spin-down energies, respectively. We do not use samples with the second, third, and sixth largest energies for the following reasons.
The object with the second-largest energy is the Crab nebula,
which is too bright for analysis without pile-up events in the soft X-ray band
\citep[for example]{tsujimoto2011};
thus, we used the previous result only with NuSTAR
\citep{madsen2015}.
The object with the third-largest energy is PSR~B0540-69, or the Crab twin in the Large Magellanic Cloud.
This pulsar experienced a spin-down rate transition in December 2011 \citep{marshall2015} and a flux change in its PWN \citep{wang2020}.
Suzaku observed this pulsar only before the transition,
whereas NuSTAR observed it after the transition.
We thus concluded that we do not make the combined spectral analysis 
with this dataset.
The object with the sixth-largest energy is PSR~J1747$-$2809, but it is not observed by NuSTAR.

The data are reprocessed with the calibration database
(ver. 20190812 for NuSTAR, ver. 20151005 for Suzaku, and  
ver. 4.8.3 for Chandra),
headas 6.24 (NusTAR, Suzaku, and Hitomi), and ciao ver. 4.9 (Chandra).
The cleaned events are extracted with the standard screening criteria
for each satellite.
The resultant exposures are listed in Table~\ref{tab:properties}.
Spectral analysis is conducted by XSPEC version 12.10.0c.

\begin{table}
  \tbl{Pulsar Properties and Observation Log.}{%
  \begin{tabular}{ccccccc}
      \hline
      Name & $\dot{E}$\footnotemark[$*$] & $\tau_c$\footnotemark[$*$] & Satellite & OBSID & Obs. start & Exposure  \\ 
      & (erg s$^{-1}$) & (yr) & & & (YYYY/MM/DD) & (ks)\\
      \hline
N157B & $4.9 \times 10^{38}$ & $4.93 \times 10^3$ & Suzaku & 806052010 & 2011/11/30 & 102 \\
  & & & NuSTAR & 40201014002 & 2016/10/17 & 105 \\
PSR~J1813$-$1749 & $5.6 \times 10^{37}$ & $5.6 \times 10^3$ & Suzaku & 401101010 & 2007/03/01 & 64 \\
  & & & NuSTAR & 30364003002 & 2018/03/25 & 27 \\
PSR~J1400$-$6325 & $5.1 \times 10^{37}$ & $1.27 \times 10^4$ & Chandra & 12567 & 2010/11/17 & 53 \\
  & & & NuSTAR & 30364001002 & 2018/05/26 & 27 \\
G21.5$-$0.9 & $3.4 \times 10^{37}$ & $4.85 \times 10^3$ & Hitomi &  100050010-100050040 & 2016/03/19 &  165(SXS)/51(SXI)/99(HXI)\footnotemark[$\dagger$]\\
      \hline
    \end{tabular}}\label{tab:properties}
\begin{tabnote}
\footnotemark[$*$] Spin-down energy ($\dot{E}$) and characteristic age ($\tau_c$) from ATNF Pulsar Catalogue v1.64 (http://www.atnf.csiro.au/research/pulsar/psrcat) \citep{manchester2005}.\\ 
\footnotemark[$\dagger$]
SXS, SXI, and HXI are the soft X-ray spectrometer \citep{kelley2016}, soft X-ray imager \citep{tanaka2018}, and hard X-ray imager \citep{nakazawa2018}, respectively.
\end{tabnote}
\end{table}

\section{Results}
\label{sec:result}

\subsection{N157B}

The coherent pulsation from the central pulsar
PSR~J0537$-$6910 was detected by \citet{marshall1998}
with the Rossi X-ray Timing Explorer (RXTE) \citep{bradt1993} and the Advanced Satellite for Cosmology and Astrophysics (ASCA) \citep{tanaka1994}.
The derived parameters (see Table~\ref{tab:properties}) show that this pulsar has the largest spin-down energy ever found in the Large Magellanic Cloud and our Galaxy.
\citet{wang2001} resolved the surrounding PWN
with Chandra,
the spectrum of which shows hardening in the region closer to the pulsar
\citep{chen2006}.
The source is also a bright gamma-ray emitter
up to very high-energy gamma-rays \citep{hess2012}.

\begin{figure}
    \centering
    \includegraphics[height=5cm]{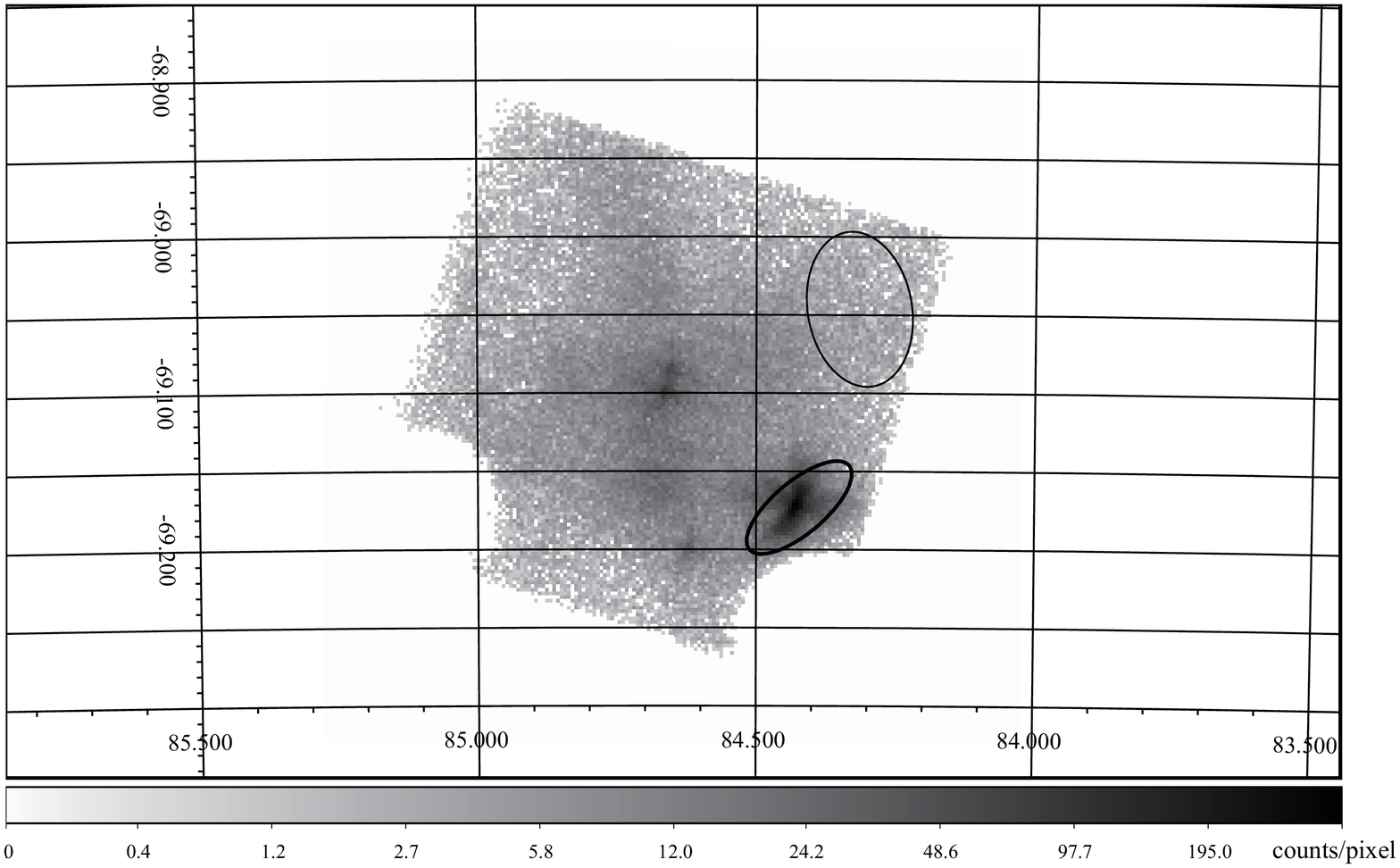}
    \includegraphics[height=5cm]{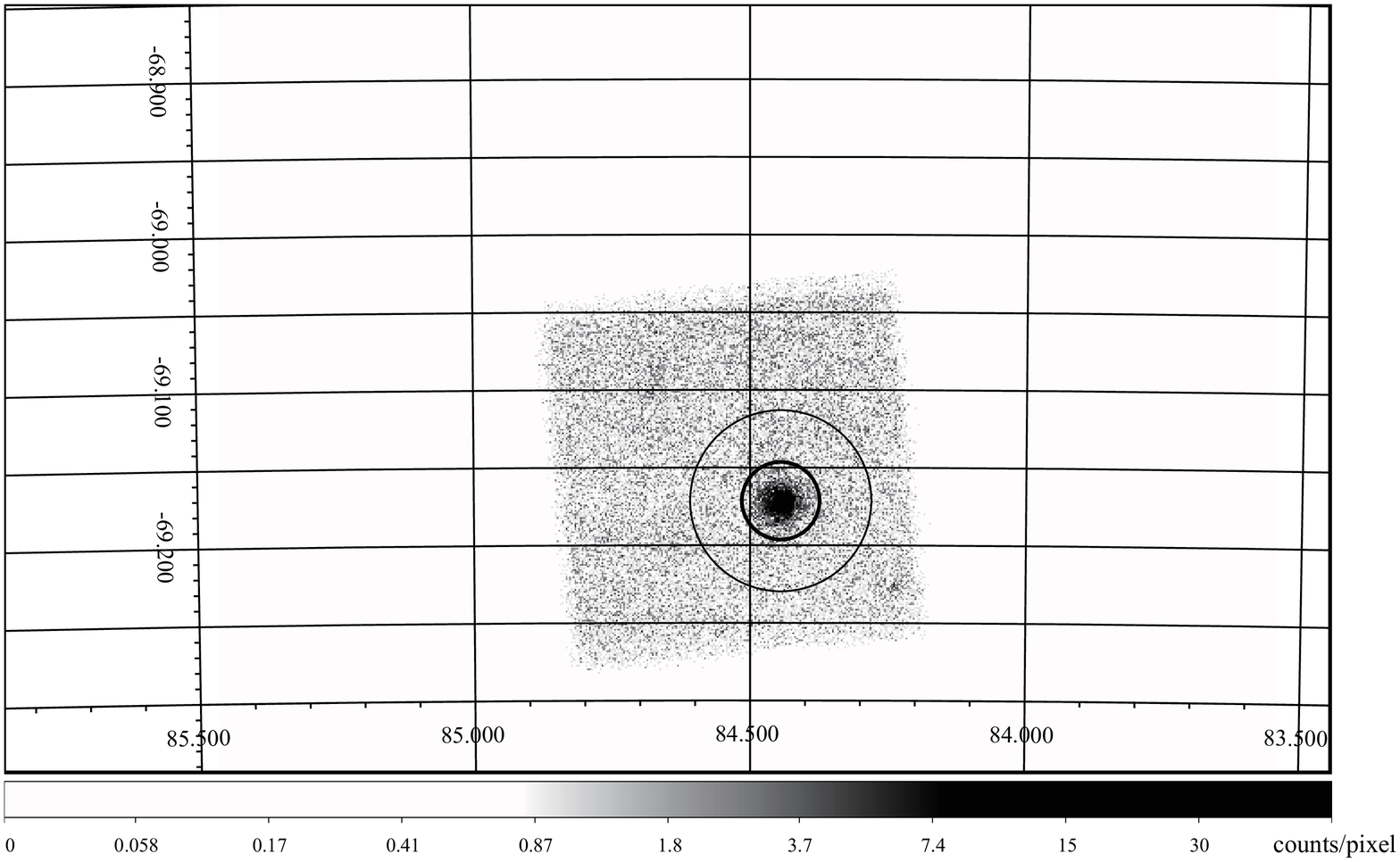}
    \caption{Suzaku XIS3 0.5--8.0~keV (left) and NuSTAR FPMA all-band (right) images of the N157B region.
    The coordinates are in J2000. The color scale is logarithmic.
    The emission in the center of the Suzaku field of view is the 30 Dor region.
    Regions enclosed by the thick and thin curves
    represent the source and background regions for the spectral analysis, respectively.}
    \label{fig:n157b-img}
\end{figure}

\begin{figure}
\begin{center}
  \includegraphics[width=8cm]{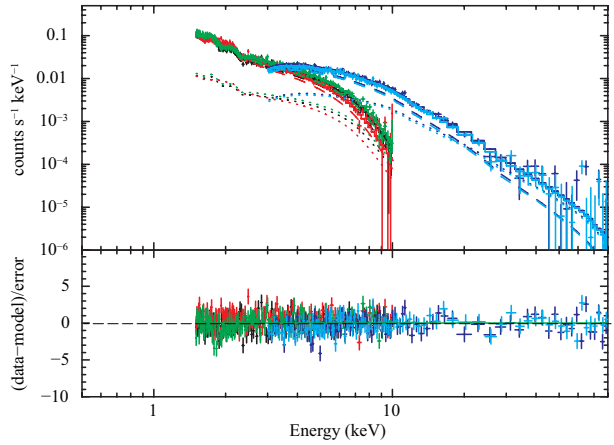} 
      \caption{Wideband spectra of N157B with the best-fit broken power-law model (solid).
      Dotted lines represent the pulsar component.
      Black, red, green, blue, and light blue symbols represent XIS0--3 onboard Suzaku, FPMA and FPMB onboard NuSTAR,
      respectively.
      The data is binned for display purposes, whereas the fit is done without binning.
     \label{fig:n157b-spec}} 
\end{center}
\end{figure}

Figure~\ref{fig:n157b-img} shows Suzaku and NuSTAR images of
the N157B region.
N157B is visible in the southwest part of the 30 Dor region,
which is located in the center of the Suzaku field of view and emits thermal X-rays \citep{cheng2021}.
Thus, we analyze the spectra above 1.5~keV
to avoid contamination from these thermal X-rays.
Figure~\ref{fig:n157b-spec} shows the spectra of the source region.
%The NuSTAR spectrum extends up to $\sim$80~keV.
We fit the spectra with the absorbed power-law, absorbed broken power-law,
and absorbed cutoff power-law models.
To represent the absorption model, we use the phabs
model in XSPEC, which includes the cross sections of \citet{balucinska-church1992} with solar abundances \citep{anders1989}.
For the pulsar component, we fix the flux and photon index derived by \citet{chen2006}.
The data are all well reproduced by these models, 
as shown in Table~\ref{tab:results}.

\subsection{PSR~J1813$-$1749}

\begin{figure}
 \begin{center}
  \includegraphics[height=5cm]{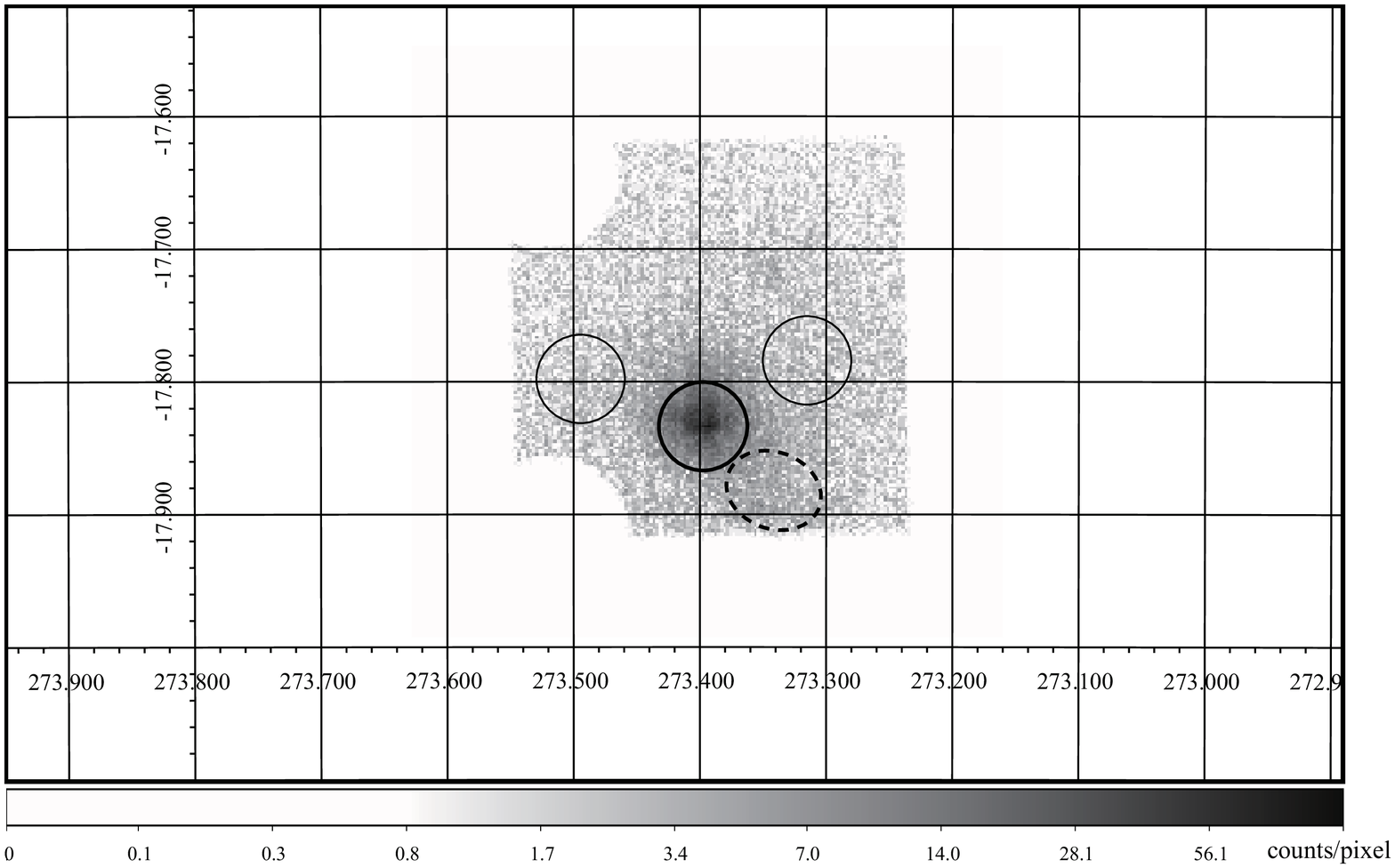} 
  \includegraphics[height=5cm]{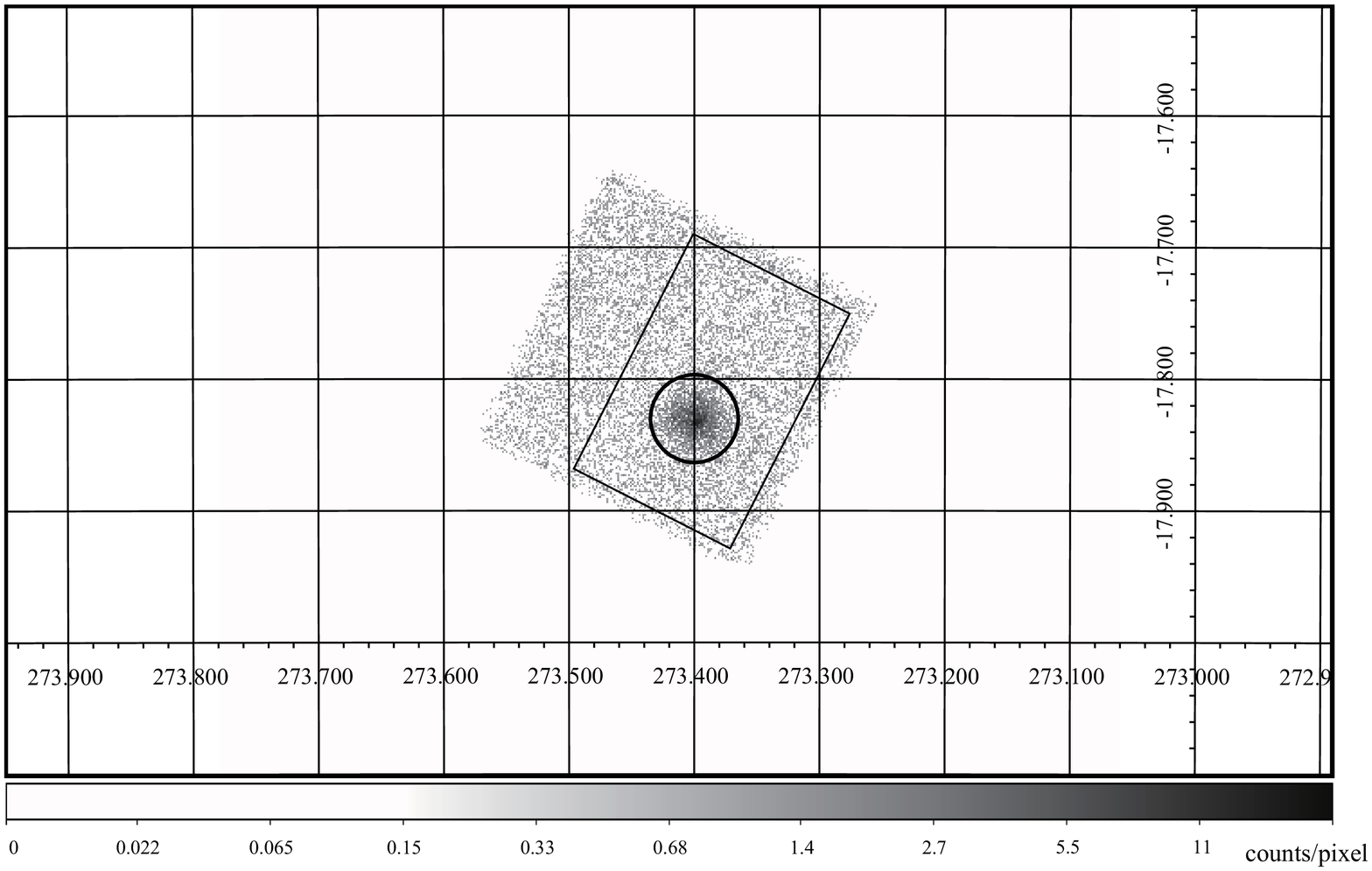} 
 \end{center}
\caption{Suzaku XIS3 0.5--8.0~keV (left) and NuSTAR FPMA all-band (right) image of the PSR~J1813$-$1749 region.
The coordinates are in J2000 and the color scale is logarithmic.
Regions enclosed by the thick and thin solid curves represent the source and background regions for the spectral analysis, respectively.
The dashed ellipse shows a soft extended emission that is possibly unrelated (see text).}\label{fig:1813-img}
\end{figure}

PSR~J1813$-$1749 was discovered as a putative pulsar with a PWN associated with the very high-energy gamma-ray source, HESS~J1813$-$178
\citep{helfand2007},
and identified as an energetic pulsar
by the discovery of its coherent pulsation \citep{gotthelf2009}.
The pulsar is inside of the young supernova remnant G12.82$-$0.02
(285--2500~yrs; \cite{brogan2005}),
which roughly agrees with the young characteristic age of the pulsar
(5600~yrs; \cite{halpern2012}).

\begin{figure}
 \begin{center}
   \includegraphics[width=8cm]{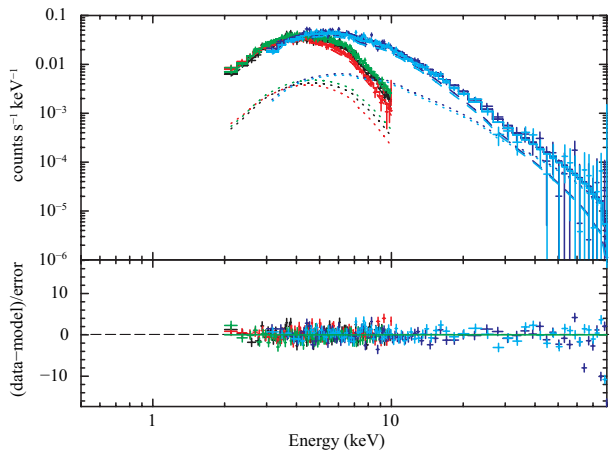} 
 \end{center}
\caption{Wideband spectra of PSR~J1813$-$1749.
The line styles and colors are the same as for N157B.
%      The data is binned for display purposes, whereas the fit is done without binning.
}\label{fig:1813-spec}
\end{figure}

Figure~\ref{fig:1813-img} shows the Suzaku and NuSTAR image of the PSR~J1813$-$1749 region.
A faint extended emission is marginally detected 
in the southwest of the pulsar 
(dashed eclipse in Fig.\ref{fig:1813-img}).
It could be a very extended component of the PWNe, as observed in several middle-aged PWNe \citep[for example]{bamba2010b}.
However, the emission is also seen below 2~keV,
which is inconsistent with the fact  
that the emissions of PSR~J1813$-$1749 are highly absorbed
\citep{brogan2005}.
\citet{brogan2005} also mentioned 
this unabsorbed component.
Therefore, we treat this emission as unrelated to our target and select the source and background regions as shown in Figure~\ref{fig:1813-img}
to avoid the extended emission.
Figure~\ref{fig:1813-spec} shows the Suzaku and NuSTAR spectra.
For the spectral analysis,
we fix the spectral parameters of the central pulsar
following \citet{helfand2007}.
Fitting is performed in the above 1.5-keV band
to avoid contamination of the emissions from the southwest of the PWN.
We use the three models in the same manner as for N157B.
The best-fit parameters are summarized in Table~\ref{tab:results}.
The large absorption is consistent with the previous results \citep{helfand2007}.
The break or cutoff of the emission just above 10~keV gives a better fit,
implying that this system is a new sample of a PWNe
with the spectral break in the X-ray band.

\subsection{PSRJ~1400$-$6325}

PSR~J1400$-$6325 was discovered by INTEGRAL
as an unidentified source \citep{keek2006},
and identified as a compact PWN by Chandra
\citep{tomsick2009}.
\citet{renaud2010} identified the central source as a pulsar from RXTE observations
by detecting a 31.18~ms pulsation.
\citet{reynolds2019} found that
some features in the pulsar-wind nebula have varied
within 6~yrs,
implying that this PWN is very energetic
like the Crab nebula \citep{camus2009}.

\begin{figure}
    \centering
    \includegraphics[height=5cm]{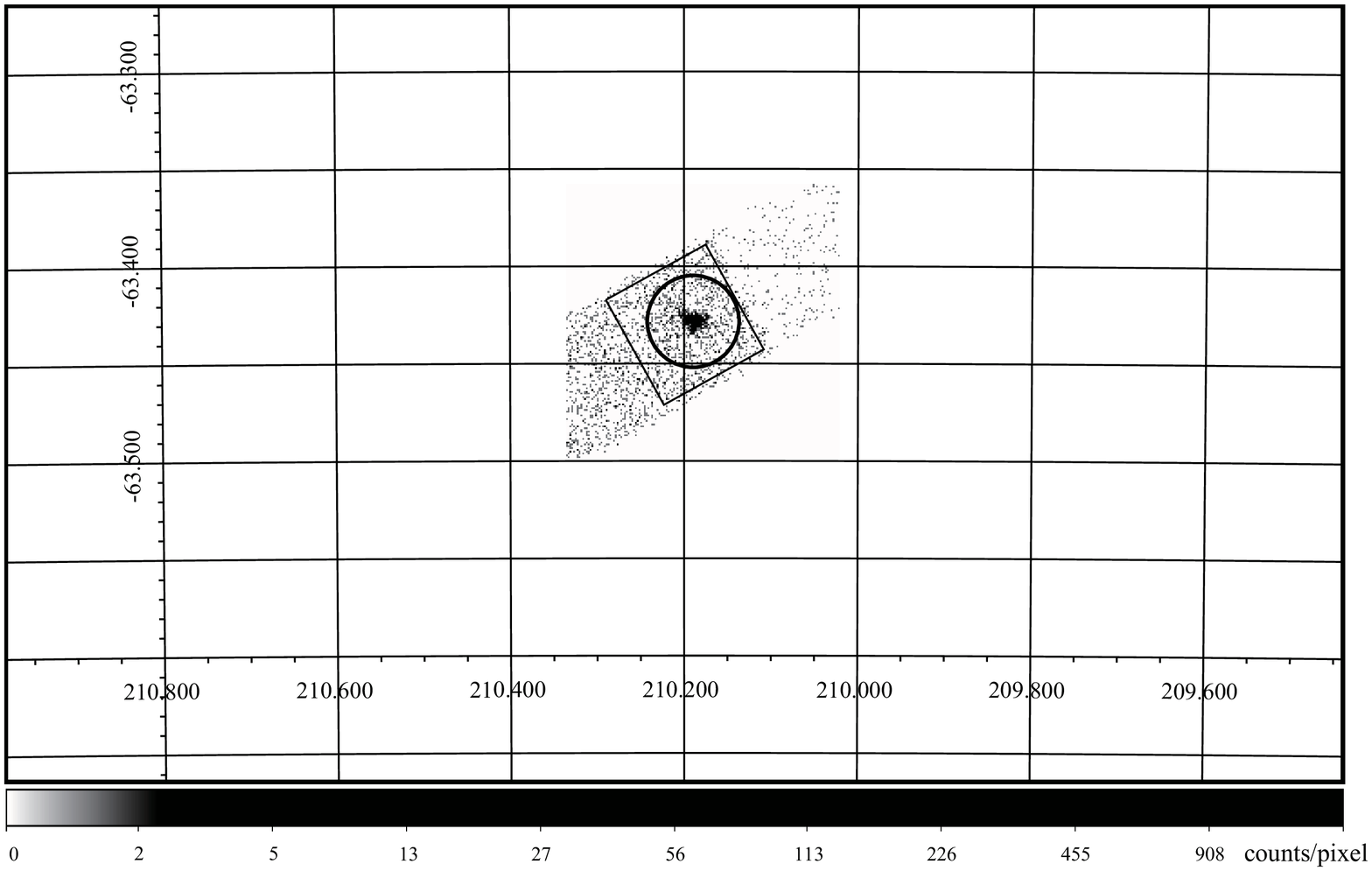}
        \includegraphics[height=5cm]{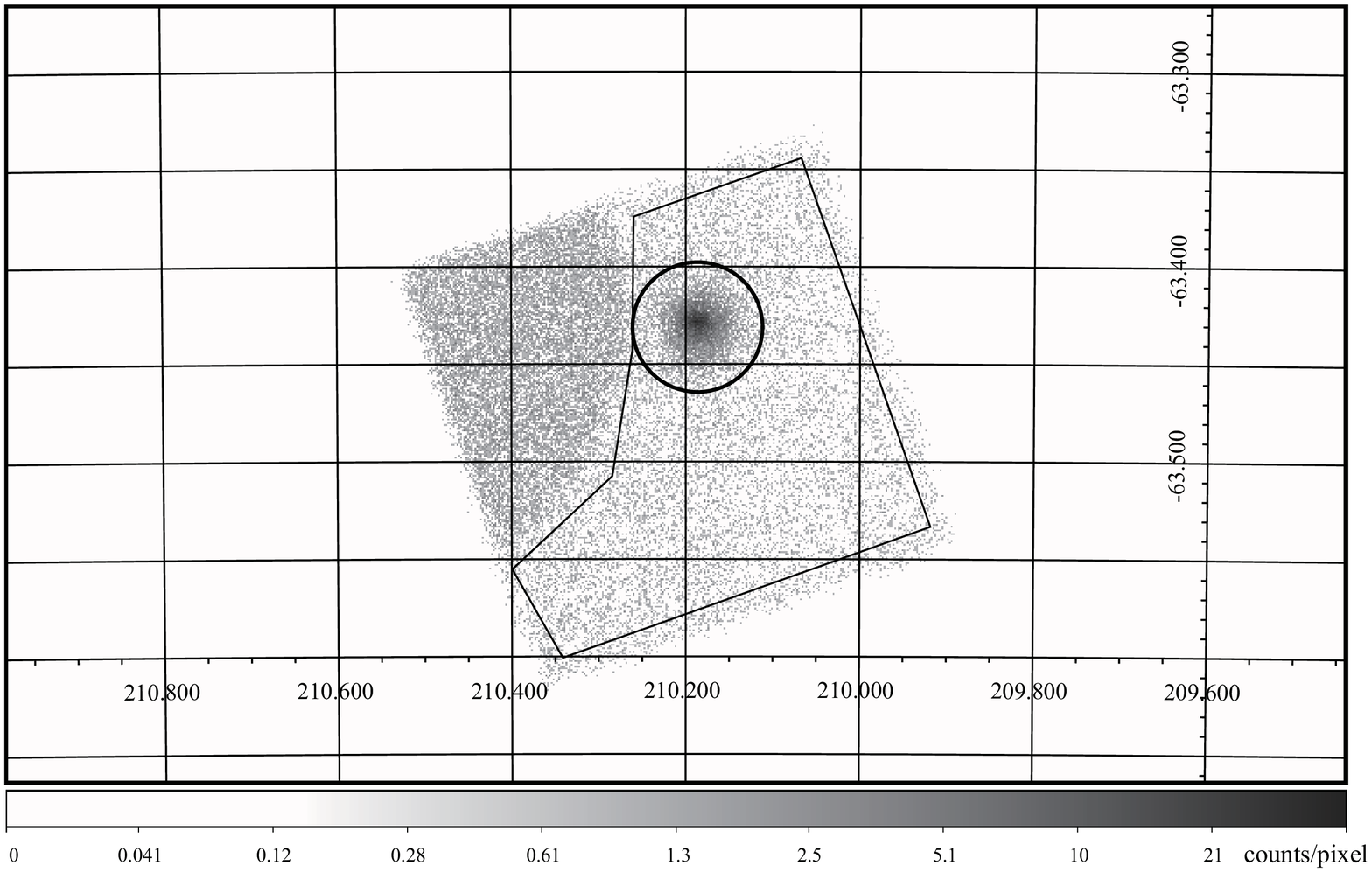}
    \caption{Chandra all-band (left) and NuSTAR FPMA all-band (right) images of PSR~J1400$-$5325.
    The coordinates are in J2000 and the color scale is logarithmic.
Regions enclosed by the thick and thin curves represent the source and background regions for the spectral analysis, respectively.
%Thick and thin regions represent source and background regions.
    The extended emission in the northeast of the NuSTAR image is due to stray light from a nearby bright source.}
    \label{fig:j1400-img}
\end{figure}

\begin{figure}
\begin{center}
      \includegraphics[width=8cm]{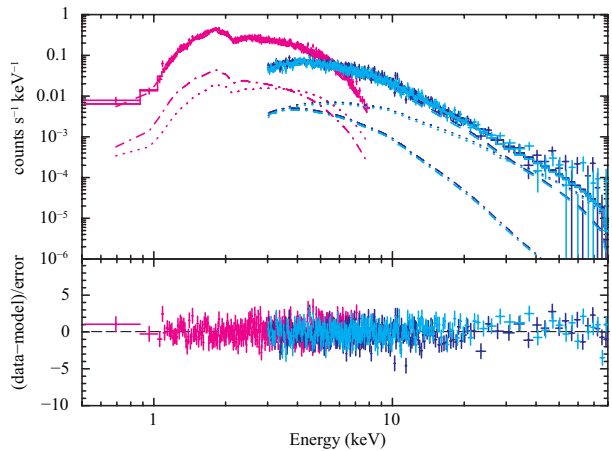} 
      \caption{PSR~J1400$-$5325 spectra and best-fit broken power-law model.
      Blue, light blue, and magenta symbols represent FPMA and FPMB onboard NuSTAR, and Chandra ACIS, respectively.
      Dotted and dashed lines represent the central pulsar and shell emissions.
      The data is binned for display purposes, whereas the fit is done without binning.
      \label{fig:j1400-spec}}
\end{center}
\end{figure}

Figure~\ref{fig:j1400-img} shows the Chandra and NuSTAR image of PSR~J1400$-$6325 region.
The PWN is surrounded by a circular shell
as shown in the Chandra image,
which emits nonthermal emissions \citep{reynolds2019}.
Because NuSTAR cannot resolve the shell spatially,
we add the component to the spectra as the fixed spectral model,
with the best-fit synchrotron radiation model (srcut) 
derived by \citet{reynolds2019}.
We also fix the spectral parameters of the central pulsar derived by \citet{renaud2010}.
The background region for the NuSTAR spectra
is selected to avoid contamination with stray light from a nearby bright source in the northeastern region outside the field of view.
Figure~\ref{fig:j1400-spec} shows the wideband spectra of the PWN.
The fitting results with the three models are summarized in Table~\ref{tab:results}.
The spectra need a clear break or cutoff, and the broken power-law model returns the best result.
The best-fit break energy is around 4~keV,
which is consistent with the previous results \citep{renaud2010}.

\subsection{G21.5$-$0.9}

G21.5$-$0.9 is one of the best observed examples of a young PWN \citep{altenhoff1970,wilson1976}
with a characteristic age of 4850~yrs \citep{gupta2005,camilo2006}.
Detailed X-ray studies with Chandra and XMM-Newton show the extended structure of the nebula
with a non-thermal power-law spectrum without line emissions below 10~keV
\citep{slane2000,safi-harb2001,warwick2001}.
\citet{tsujimoto2011} found that the photon index of the nebula above 10~keV is larger than that below 10~keV,
in their cross-calibration processes,
and \citet{nynka2014} confirmed it with NuSTAR.
\citet{hitomi2018} performed wideband spectroscopy 
and measured the break energy precisely, 
$7.1\pm0.3$~keV.

We use 
exactly the same unbinned spectra as \citet{hitomi2018}
to derive the results with the cutoff power-law model.
Following the analysis by \citet{hitomi2018},
we define the model "composite" as multiple components from the pulsar, the extended halo and the limb, a weak, thermal soft component from the northern knot, is represented by a non-equilibrium ionization model ({\tt vpshock} in xspec)
\citep{matheson2010,guest2019,bocchino2005}.
The fitting results are summarized in Table~\ref{tab:results}.
%The fitting returned almost the same c-statistics to that of the broken power-law model fitting \citep{hitomi2018}.

\subsection{Fitting Summary}

The fitting results for the four targets are 
summarized in Table~\ref{tab:results}.
The C-statistics values for each target are smaller in the broken power-law or cutoff power-law models
compared with that for a simple power-law model.
To estimate how significant it is that the spectra need a break or cutoff,
we derive a $p$-value via a likelihood ratio test.
The results are also shown in Table~\ref{tab:results}.
The power-law model is rejected with $p\ll 0.01$
and the data require a significant break or cutoff for all the samples except for N157B.
The $p$-value for N157B is around 10\%,
implying the break or cutoff is not significant
in this case.

\begin{figure}
    \centering
    \includegraphics[width=0.4\textwidth]{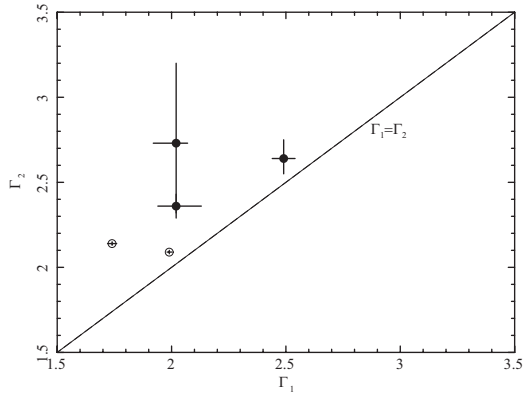}
    \caption{Photon indices below ($\Gamma_1$) and above ($\Gamma_2$) in the broken power-law fitting. The filled data points indicate our results and the open data points indicate results from previous studies (the Crab and G21.5$-$0.9 results from NuSTAR and Hitomi \citep{madsen2015,hitomi2018}). The dashed line shows $\Gamma_1 =  \Gamma_2$.
    The errors are the 90\% confidence level.}
    \label{fig:gamma1-gamma2}
\end{figure}

Figure~\ref{fig:gamma1-gamma2} compares the photon index
below and above the break
in the broken power-law model fitting.
We add the Crab results from NuSTAR,
which show the spectral change from $\Gamma = 1.99\pm0.01$ to $2.09\pm0.01$ at a break energy of $8.3\pm0.2$~keV \citep{madsen2015}
in addition to our sample.
All targets require a larger photon index in the harder energy band,
which also suggests that our targets need a spectral break or cutoff to reproduce the wideband spectra.

\begin{table}
  \tbl{Fitting Summary.\footnotemark[$*$]}{%
  \begin{tabular}{lcccc}
      \hline
      & N157B & PSR~J1813$-$1749 & PSR~J1400$-$6325 & G21.5$-$0.9 \\
      \hline
      \multicolumn{5}{c}{Absorbed power-law model}\\
      \hline
      $N_{\rm H}$ ($10^{22}$~cm$^{-2}$) & 0.58 (0.52--0.64) & 10.1 (9.8--10.4) & 2.63 (2.59--2.68) & --- \\
      $\Gamma$ & 2.534 (2.51--2.57) & 2.11 (2.08--2.15) & 2.28 (2.26--2.30) & --- \\
      $F_{\rm 2-10~keV}$ ($10^{-12}$erg~cm$^{-2}$s$^{-1}$)\footnotemark[$\dagger$] & 4.11 & 7.77 & 1.32 & --- \\
      C-statistics/d.o.f. & 11872.3/10829 & 11110.6/10418 & 4786.8/4357 & --- \\
      \hline
      \multicolumn{5}{c}{Absorbed broken power-law model}\\
      \hline
      $N_{\rm H}$ ($10^{22}$~cm$^{-2}$) & 0.52 (0.43--0.60) &  9.7 (9.4--10.0) & 2.39 (2.32--2.49) &  --- \\
      $\Gamma_1$ & 2.49 (2.44--2.54) &  2.02 (1.92--2.07) & 2.02 (1.94--2.13) & --- \\
      $E_{\rm br}$ (keV) & 7.4 (5.2--11.4) & 14.0 (8.1--16.6) & 4.2 (3.9--5.2) & --- \\
      $\Gamma_2$ & 2.64 (2.55--2.75) & 2.73 (2.29--3.20) & 2.36 (2.32--2.43) & --- \\
      $F_{\rm 2-10~keV}$ ($10^{-12}$erg~cm$^{-2}$s$^{-1}$)\footnotemark[$\dagger$]  & 4.11 & 7.81 & 1.32 & --- \\
      C-statistics/d.o.f. &  11868.2/10827 & 11077.54/10416 & 4737.2/4355 & --- \\
      $p$-value\footnotemark[$\ddagger$] & 0.129 & $6.62\times 10^{-8}$ & $1.70\times 10^{-11}$ & $5.50\times 10^{-251}$\footnotemark[\S] \\
      \hline
            \multicolumn{5}{c}{Absorbed cutoff power-law model}\\
      \hline
      $N_{\rm H}$ ($10^{22}$~cm$^{-2}$) &  0.50 (0.40--0.60) & 9.0 (8.6--9.4) & 2.54 (2.48--2.60) & 3.29 (3.26--3.32) \\
      $\Gamma$ & 2.44 (2.32--2.54) & 1.68 (1.54--1.81) & 2.13 (2.05--2.21) & 1.70 (1.68--1.72) \\
      $E_{\rm cut}$ (keV) & 67.3 ($>$ 32.6) & 22.1 (16.9--31.2) & 46.0 (30.0--94.1) & 34.3 (32.4--36.4) \\
      $F_{\rm 2-10~keV}$ ($10^{-12}$erg~cm$^{-2}$s$^{-1})$\footnotemark[$\dagger$]  & 4.10 & 7.81 & 1.32 &  47.0 \\
      C-statistics/d.o.f. & 11869.6/10828 & 11072.1/10417 & 4775.9/4356 & 24429.8/23028 \\
    $p$-value\footnotemark[$\ddagger$] & 0.100 & $5.48\times 10^{-10}$ & $9.62\times 10^{-4}$ & $8.58\times 10^{209}$\footnotemark[\S] \\
      \hline
      \end{tabular}}\label{tab:results}
\begin{tabnote}
\footnotemark[$*$] The errors are the 90\% confidence level.\\ 
\footnotemark[$\dagger$] Observed flux in the 2--10~keV band.\\
\footnotemark[$\ddagger$] The probability that the power-law model represents the spectra better than the broken/cutoff power-law models.\\
\footnotemark[\S] Derived from the comparison of C-statistics in \citet{hitomi2018}.
\end{tabnote}
\end{table}

\section{Comparison of Spectral Parameters}
\label{sec:comparison}

\begin{figure}
     \centering
       \includegraphics[width=0.42\textwidth]{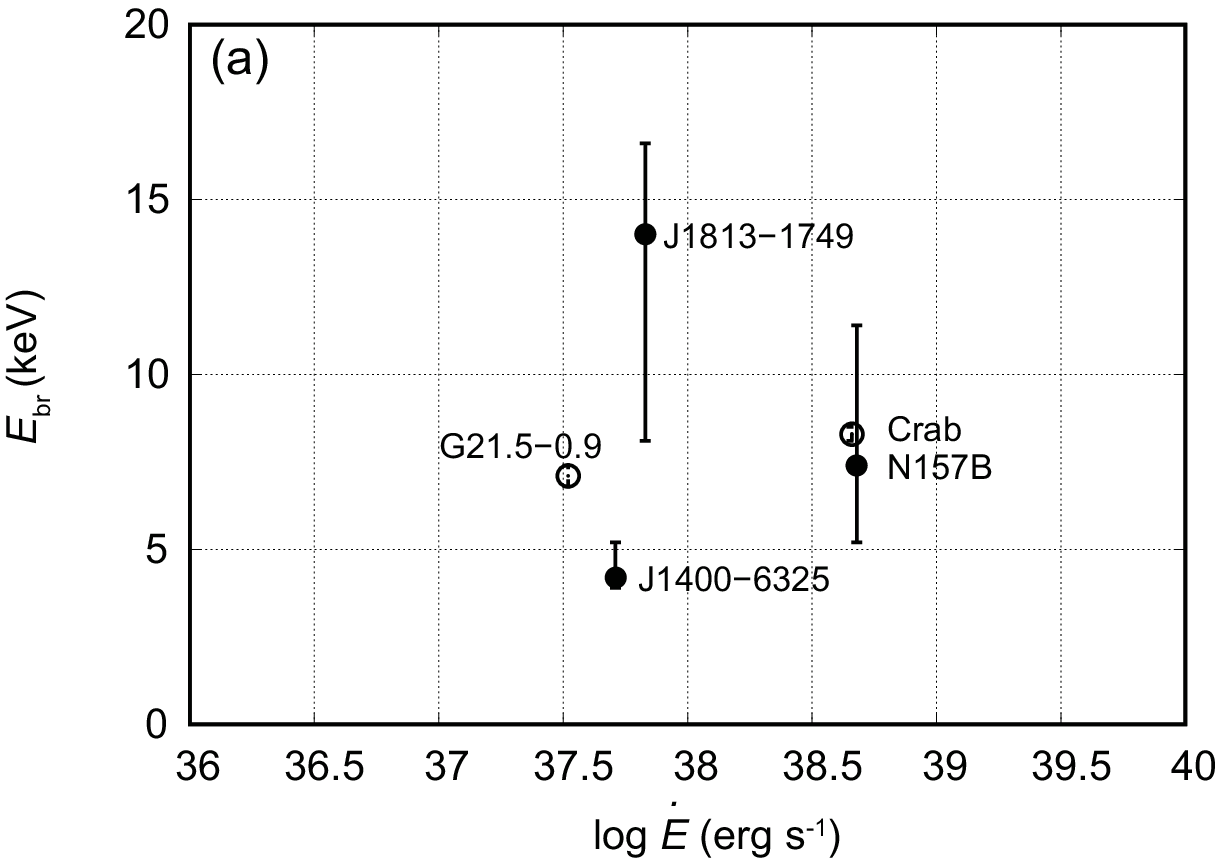}
       \includegraphics[width=0.42\textwidth]{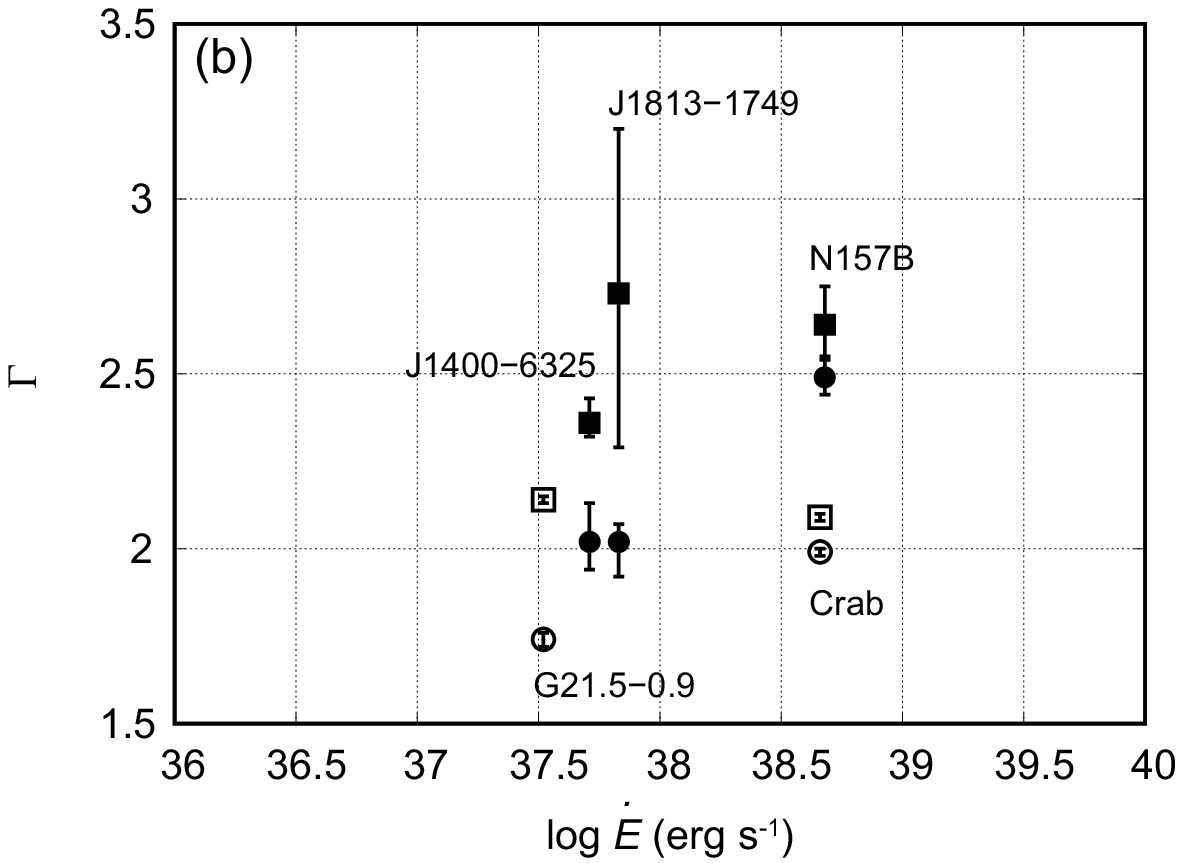}
       \includegraphics[width=0.42\textwidth]{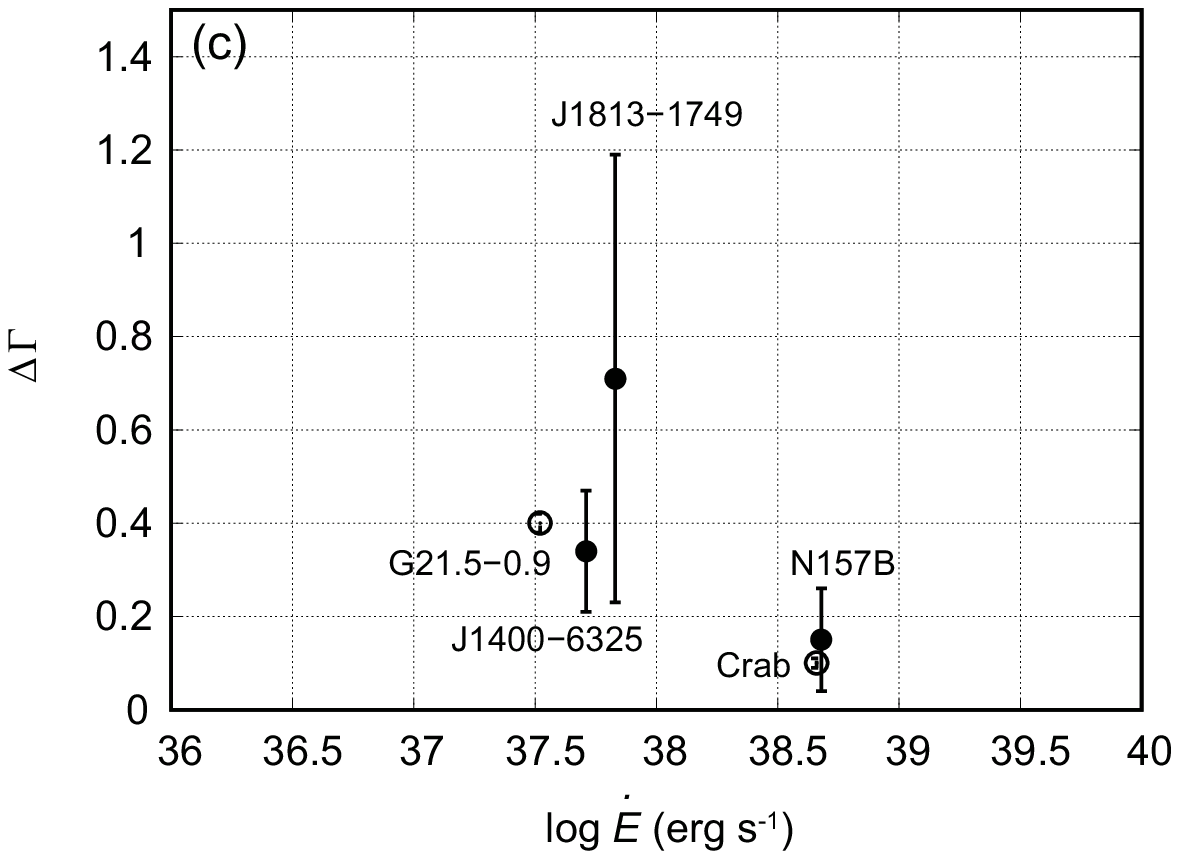}
       \includegraphics[width=0.42\textwidth]{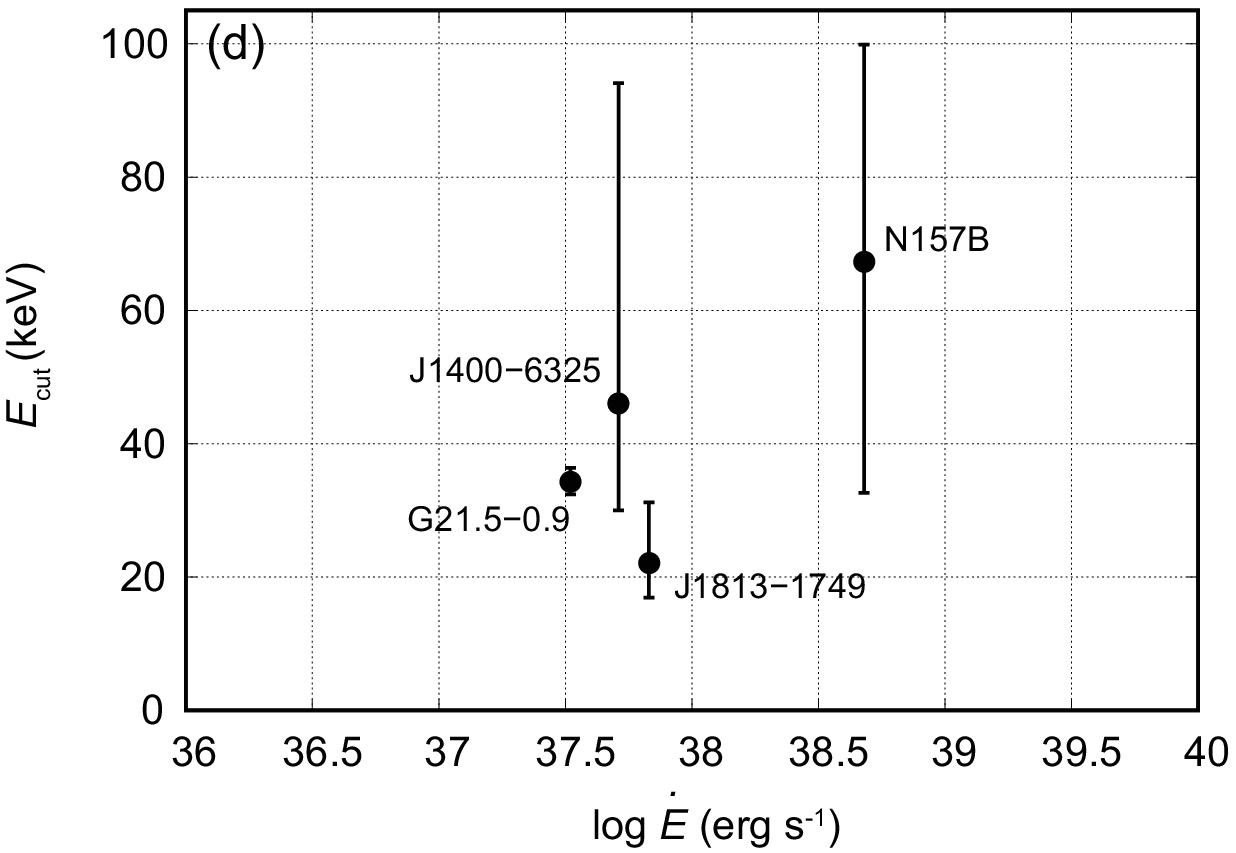}
       \includegraphics[width=0.42\textwidth]{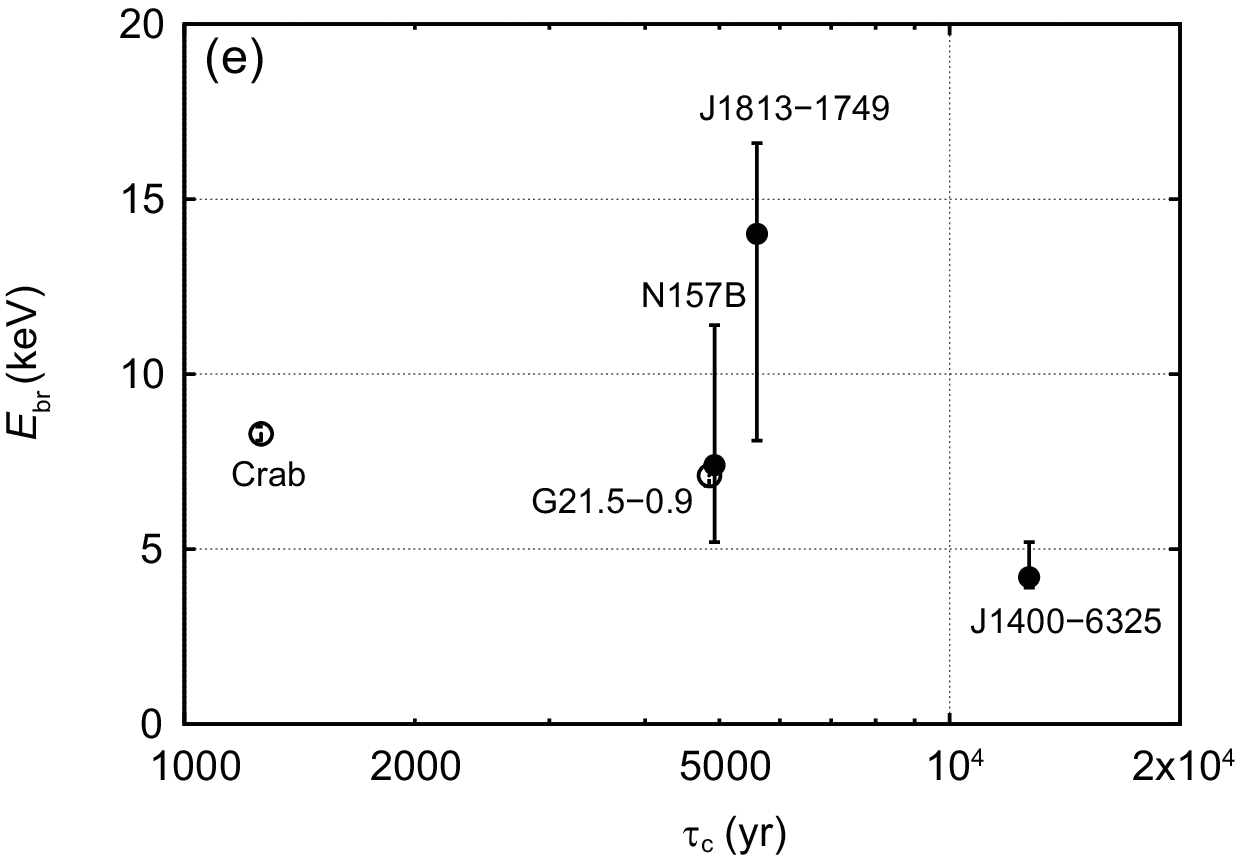}
       \includegraphics[width=0.42\textwidth]{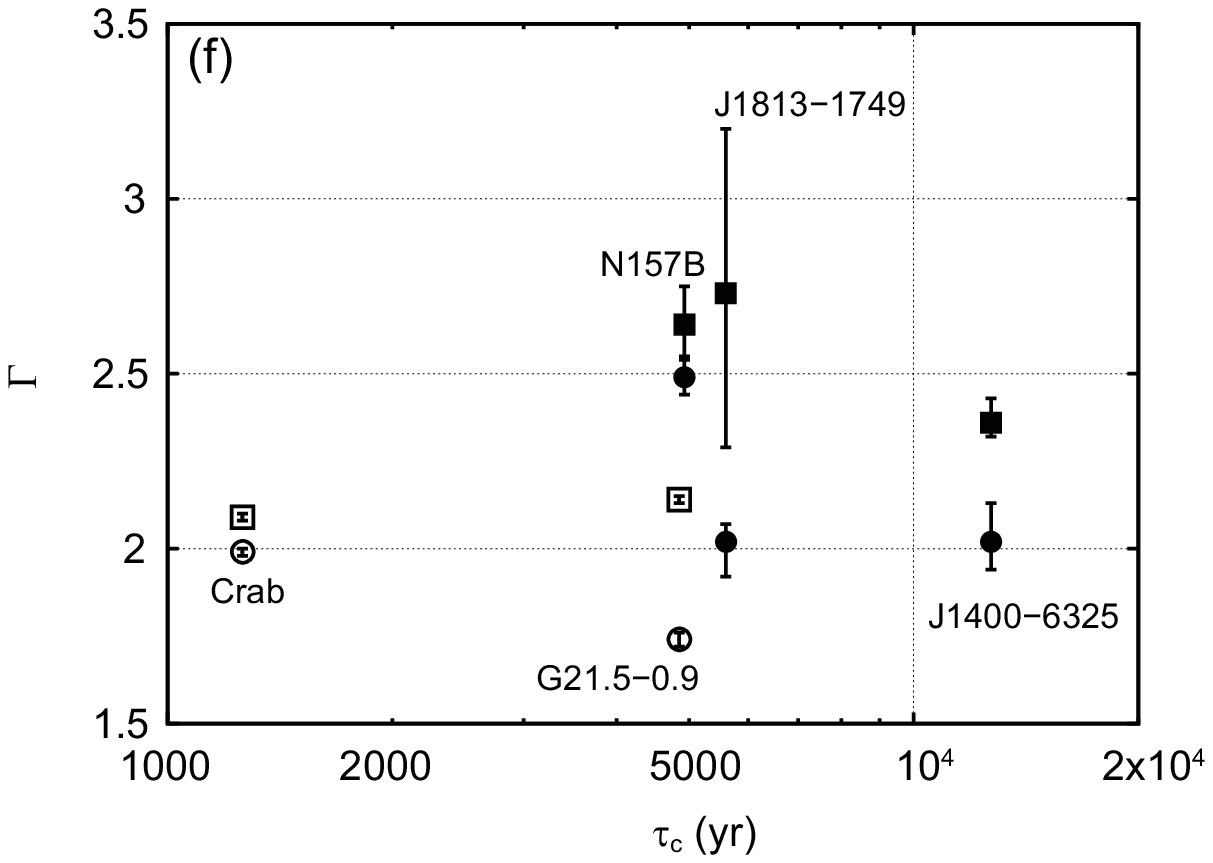}       \includegraphics[width=0.42\textwidth]{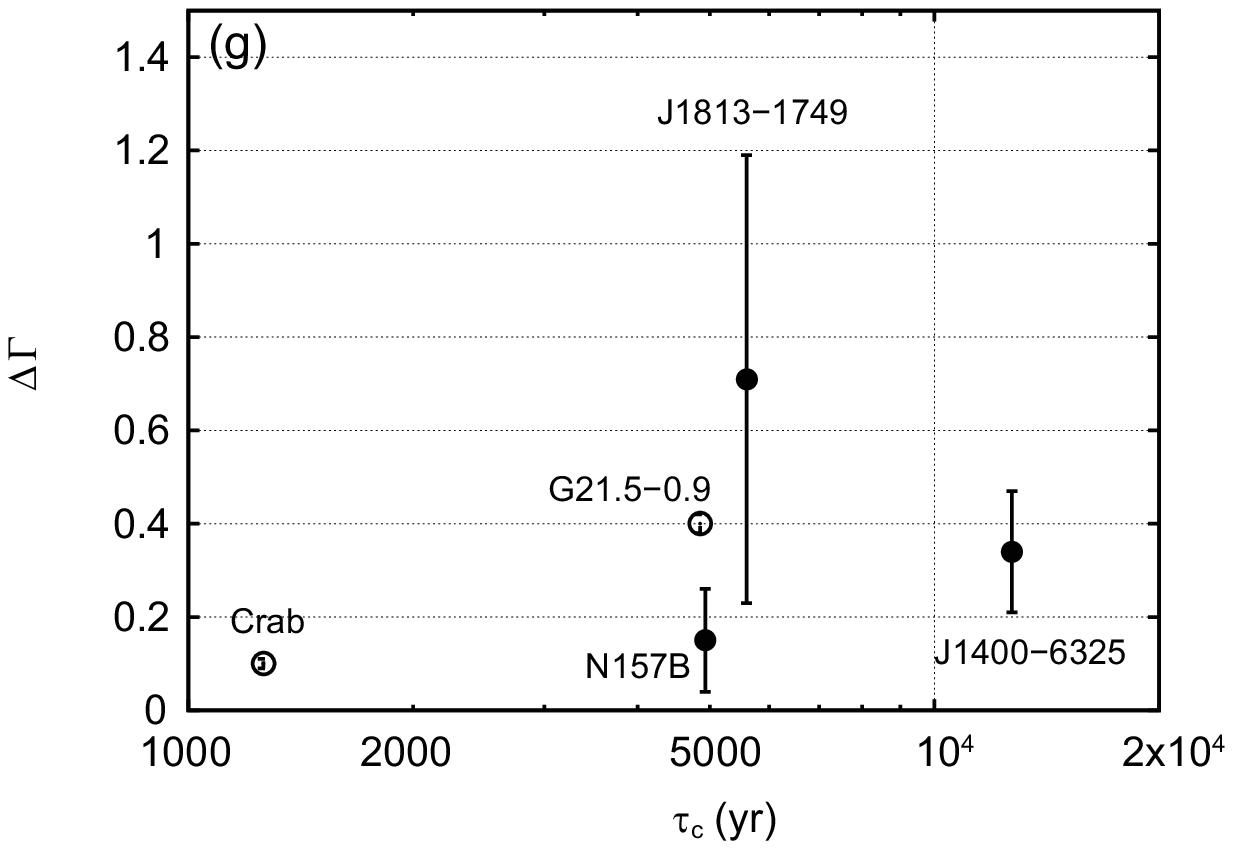}
       \includegraphics[width=0.42\textwidth]{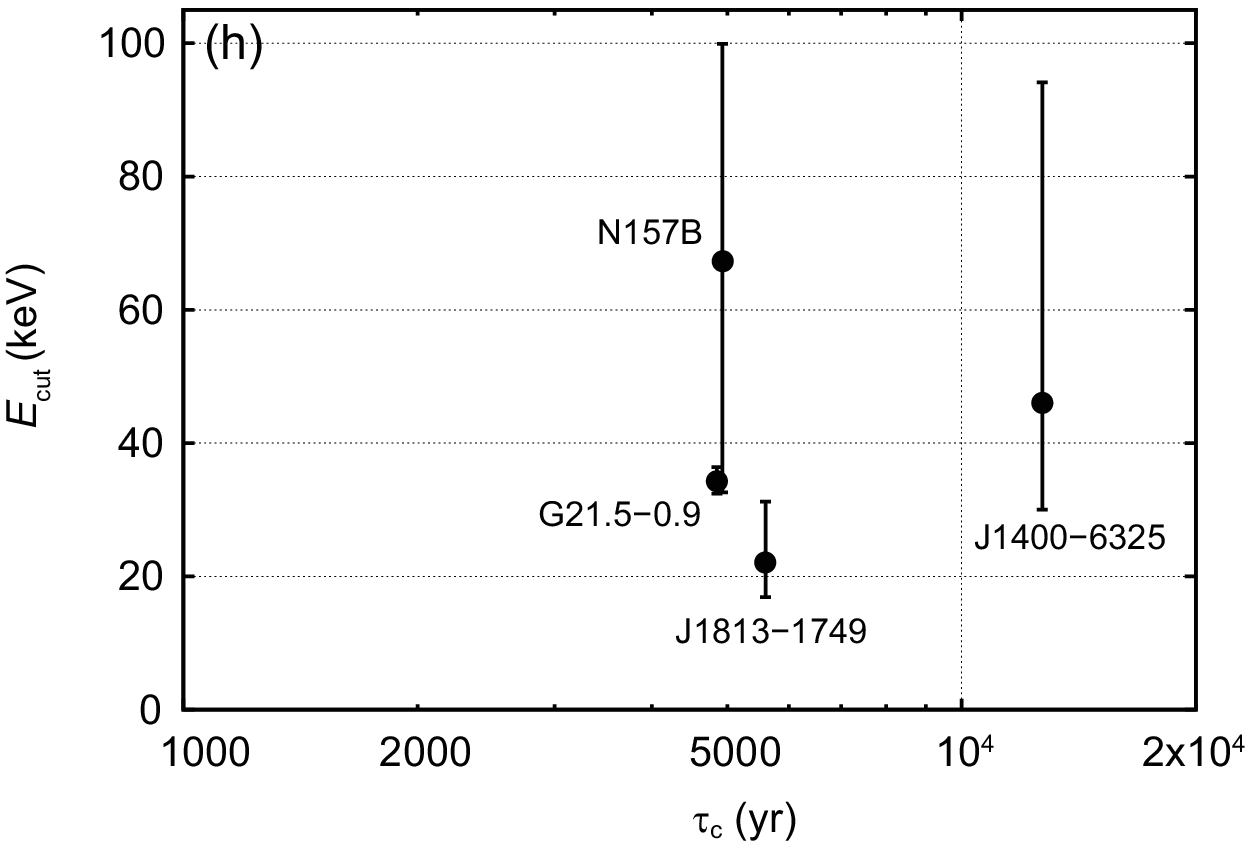}
    \caption{Comparison of pulsar and derived parameters. Filled data points show our results and open points are taken from literature \citep{madsen2015,hitomi2018}.
    In panels (b) and (f), circles indicate $\Gamma_1$ data and boxes indicate $\Gamma_2$ data.
    The error represents the 90\% confidence range.}
     \label{fig:comparison}
 \end{figure}

The wideband spectrum of PWNe has a significant break or cutoff for three of our four samples.
In this section, we compare these spectral parameters and pulsar properties.
Many observational parameters have a correlation between their spin-down energy or characteristic age \citep[for example]{kargaltsev2008}; thus, we first 
plot our results with these two parameters.
Figure~\ref{fig:comparison} shows our derived parameters, which are $E_{\rm br}$ from the broken power-law fitting,
$\Gamma_1$ before the break, $\Gamma_2$ after the break,
$\Delta\Gamma \equiv \Gamma_2 - \Gamma_1$, and $E_{\rm cut}$ derived from the cutoff power-law model fitting, plotted versus the spin-down energy ($\dot{E}$).
The first and second columns of Table~\ref{tab:correlation} summarize the correlation coefficients of pulsar parameters $\log\dot{E}$ and $\log\tau_c$, and X-ray spectral parameters.
The error range of the correlation coefficients are derived considering the errors of spectral fitting.
The absolute value of correlation coefficients must be larger than 0.81 
to demonstrate correlation within the 90\% confidence level for the sample number of five;
thus, none of the spectral parameters is correlated with $\log\dot{E}$ and $\log\tau_c$
significantly.
Among these possible relationships, $\log\dot{E}$ and $\Gamma_1$ may have some degree of correlation because they have relatively large correlation coefficients.
Considering that $\Gamma_1$ in the broken power-law model is the photon index below 10~keV,
the possible correlation between $\log\dot{E}$ and $\Gamma_1$ has already been reported by \citet{li2008},
and they discussed this tendency as indicating more cooling with larger $\dot{E}$ samples
with the models by \citet{chevalier2000}.
Our results indicate that the tendency remains, even if we consider the spectral break and the derived $\Gamma_1$ to be the value without affecting cooling; 
this fact implies that another scenario is required.

\begin{table}
  \tbl{Correlation Coefficients for Pulsar Parameters and Spectral Parameters.\footnotemark[$*$]}{%
  \begin{tabular}{p{3pc}ccc}
      \hline
       & $\log \dot{E}$ & $\log\tau_c$  & $\log\eta_{\rm PSR}$ \\ 
      \hline
      $E_{\rm br}$\dotfill & $-$0.03 ($-$0.20 -- 0.53) & $-$0.27 ($-$0.70 -- $-$0.14) & $-$0.37 ($-$0.49 -- 0.34)  \\
      $\Gamma_1$\dotfill & 0.70 (0.63 -- 0.75) & 0.06 ($-$0.03 -- 0.18) & 0.26 (0.18 -- 0.37) \\
      $\Gamma_2$\dotfill & 0.03 ($-$0.15 -- 0.33) & 0.45 (0.33 -- 0.56) & $-$0.54 ($-$0.69 -- $-$0.07) \\
      $\Delta\Gamma$\dotfill & $-$0.75 ($-$0.96 -- $-$0.49) &  0.46 (0.23 -- 0.77) & $-$0.93 ($-$0.97 -- $-$0.50) \\
      $E_{\rm cut}$\dotfill & 0.83 ($-$0.20 -- 0.96) & 0.04 ($-$0.37 -- 0.89) & 0.87 (0.04 -- 0.94) \\
      \hline
    \end{tabular}}\label{tab:correlation}
\begin{tabnote}
\footnotemark[$*$] The errors are the 90\% confidence level.\\  
\end{tabnote}
\end{table}

\section{Discussion}
\label{sec:discuss}

We find significant X-ray spectral curvature in the emissions from all the PWN samples, except for N157B.
In this section, we try to reproduce this curvature
with simple modelling.

\subsection{Mechanisms of X-Ray Spectral Curvature}

X-ray emissions from PWNe originate from the non-thermal particles injected with a standard single power-law distribution \citep{kennel1984b}.
One way to create the observed convex spectral curvature in the X-ray emissions is a spectral cutoff and the other is a spectral break.
Here, we discuss typical photon energies for some spectral cutoff and break mechanisms and compare them with the observed values because the observed X-ray spectral curvatures are reproduced well both by the absorbed cutoff and broken power-law models.
There is much room for discussion on the spectral cutoff
and break until observations are available at higher energies than the present ones.
%Note that the spectral cutoff and break can be distinguished by observing higher energy photons than the present analysis.

In the one-zone approximation, the energy distribution of the electrons, $N(\gamma, t)$, follows
\begin{equation}\label{distribution}
    \frac{\partial N}{\partial t}
    +
    \frac{\partial}{\partial \gamma} (\dot{\gamma}_{\rm cool} N)
    =
    Q_{\rm inj}
    -
    \frac{N}{t_{\rm esc}},
\end{equation}
where $\gamma$ is the Lorentz factor of electrons.
We consider synchrotron cooling as the dominant cooling process, $\dot{\gamma}_{\rm cool} = \dot{\gamma}_{\rm syn}$, and the diffusive escape of the time-scale, $t_{\rm esc}$.
The injection of accelerated particles $Q_{\rm inj}$ is set to the single power-law distribution $Q_{\rm inj} \propto  \gamma^{-p}$ with a cutoff energy of $\gamma_{\rm max}$ \citep{blandford1987}.
Following the discussion by \citet{tanaka2010}, we obtain the approximated expression of $N(\gamma, t)$ in Equation (\ref{distribution}) as
\begin{equation}\label{approx}
    N(\gamma, t)
    \sim
    \left(
    t^{-1}_{\rm age}
    +
    t^{-1}_{\rm syn}(\gamma, t)
    +
    t^{-1}_{\rm esc}(\gamma, t)
    \right)^{-1}
    Q_{\rm inj}(\gamma, t),
\end{equation}
where we use the approximations that $\partial / \partial t \sim t_{\rm age}^{-1}$, where $t_{\rm age}$ is the age of the system, and $\partial \dot{\gamma}_{\rm syn}/\partial \gamma \sim t_{\rm syn}^{-1}$.

The observed spectral cutoff could correspond to the synchrotron photons emitted by the maximum energy of the accelerated particles, $\gamma_{\rm max}$.
Considering the particle acceleration at the termination shock of the pulsar wind, $\gamma_{\rm max}$ can be estimated by equating the gyro-radius of the particles and the size of the termination shock, $r_{\rm TS}$ \citep{kennel1984b}.
For a given magnetic field strength inside the PWN, $B_{\rm PWN}$, corresponding cutoff synchrotron photon energy $\varepsilon_{\rm size}$ is
\begin{equation}
    \varepsilon_{\rm size}
    =
    164~{\rm keV} \left(\frac{B_{\rm PWN}}{10 ~\mu G}\right)^3 \left(\frac{r_{\rm TS}}{0.1 ~{\rm pc}}\right)^2.
\end{equation}
% Another estimate of $\gamma_{\rm max}$ can be obtained from the polar cap potential of the central pulsar \citep[for example]{bucciantini2011}.
% % \begin{equation}
% %     \gamma_{\rm w} \kappa (1 + \sigma_{\rm w}) \approx 1.4 \times 10^{10} \left( \frac{L_{\rm spin}}{10^{38}~{\rm erg~ s^{-1}}} \right)^{1/2},
% % \end{equation}
% % where $\gamma_{\rm w}$, $\kappa$ and $\sigma_{\rm w}$ are the bulk Lorentz factor, pair multiplicity and magnetization of the pulsar wind.
% Corresponding cutoff synchrotron photon energy {\bf $\varepsilon_{\rm PC}$} is
% \begin{equation}
%     \varepsilon_{\rm PC}
%     =
%     98.6~{\rm keV} \left(\frac{B_{\rm PWN}}{10 ~\mu G}\right) \left(\frac{E_{\rm dot}}{10^{37}~{\rm erg~s^{-1}}}\right).
% %    9.86 \times 10^{-1}~{\rm eV} \left(\frac{B_{\rm PWN}}{10 ~\mu G}\right) \left(\frac{L_{\rm spin}}{10^{38}~{\rm erg~s^{-1}}}\right) \left(\frac{\kappa}{10^{3}}\right)^{-2} (1+\sigma)^{-2}.
% \end{equation}
% $\varepsilon_{\rm size}$ and $\varepsilon_{\rm PC}$ for each object are summarized in Table \ref{tbl:TypicalPhotonEnergy}, where $B_{\rm PWN}$ and $r_{\rm TS}$ are adopted from \citet{zhu2018}.
% Both of them are significantly higher than the obtained values of the cutoff photon energy in the present analysis.
The values of $\varepsilon_{\rm size}$ for each object are summarized in Table \ref{tbl:TypicalPhotonEnergy}, where $B_{\rm PWN}$ and $r_{\rm TS}$ are adopted from \citet{zhu2018}.
We conclude that $\varepsilon_{\rm size}$ is too high to explain the cutoff photon energy, $E_{\rm cut}$, in the present analysis.
A comprehensive study of broadband spectra of PWNe by \citet{zhu2018} modeled all the objects in the present analysis at once.
Although the obtained values of $B_{\rm PWN}$ for each object are close but slightly different from other studies of broadband modeling of PWNe, the relative values of $B_{\rm PWN}$ between objects are similar from model to model \citep[for example]{tanaka2010,tanaka2011,tanaka2013,bucciantini2011,torres2014}.

Next, we discuss the spectral break mechanisms.
The typical electron energy corresponding to synchrotron cooling break $\gamma_{\rm syn}$ is given by equating the synchrotron cooling time, $t_{\rm syn}(\gamma)$, with the age of the system, $t_{\rm age}$, and the corresponding synchrotron photon energy, $\varepsilon_{\rm syn}$, is
\begin{equation}\label{eq:SynBreakEnergy}
    \varepsilon_{\rm syn}
    =
    3.0~{\rm keV}
    \left( \frac{B_{\rm PWN}}{10 \mu{\rm G}} \right)^{-3}
    \left( \frac{t_{\rm age}}{1 {\rm kyr}  } \right)^{-2}.
\end{equation}
The distribution function above $\gamma_{\rm syn} (\ll \gamma_{\rm max})$ can be approximated as $N(\gamma > \gamma_{\rm syn}) \approx Q_{\rm inj} t_{\rm syn}(\gamma) \propto \gamma^{-p - 1}$ by setting $t_{\rm syn} \ll t_{\rm age}, t_{\rm esc}$ in Equation (\ref{approx}).
The distribution function below $\gamma_{\rm syn}$ is obtained as $N(\gamma < \gamma_{\rm syn}) \approx Q_{\rm inj} t_{\rm age} \propto \gamma^{-p}$ by setting $t_{\rm age} \ll t_{\rm syn}, t_{\rm esc}$ in Equation (\ref{approx}).
We obtain a rough estimate of the amount of the spectral break as
\begin{equation}\label{eq:SynBreakSize}
	\Delta\Gamma_{\rm syn} = 0.5.
\end{equation}
Equations (\ref{eq:SynBreakEnergy}) and (\ref{eq:SynBreakSize}) are the characteristics of the synchrotron cooling break in the photon spectrum.

The particle escape process remains unknown; thus, we discuss the three diffusion processes here.
The first is the Bohm diffusion
\begin{equation}
    D_{\rm Bohm}(\gamma)
    =
    \frac{\zeta}{3} r_{\rm g} c,
\end{equation}
where we introduce the Bohm factor $\zeta (>1)$ and gyro-radius $r_{\rm g} = \gamma m_{\rm e} c^2 / e B_{\rm PWN}$,
where $m_{\rm e}$ and $c$ are the electron mass and the light velocity, respectively.
For the second and third processes, we consider the resonant scattering by the turbulent magnetic field along or across the background magnetic field \citep[for example]{blandford1987},
\begin{equation}
    D_{\parallel}(\gamma) \approx \frac{v^2}{3 \nu_{{\rm res},k}}
    ,~{\rm and}~
    D_{\perp}(\gamma)     \approx \frac{v^2 \nu_{{\rm res},k}}{3 \omega^2_{\rm g}},
\end{equation}
where $v$ is the particle velocity and $\omega_{\rm g} = e B_{\rm PWN} / \gamma m_{\rm e} c$ is the gyro-frequency.
The resonant scattering frequency, $\nu_{{\rm res},k} = (\pi/4) (8 \pi k \epsilon_k/B^2_{\rm PWN}) \omega_{\rm g}$, can be calculated when we specify the spectrum of the magnetic turbulence, $\epsilon_k \propto k^{-q}$, and hereafter the Kolmogorov type turbulence $q = 5/3$ is adopted \citep[for example]{porth2016}.
We introduce parameter $\xi$, which gives the energy density of the turbulent magnetic field relative to the background magnetic field, $\int^{\infty}_{k_{\rm min}} \epsilon_k d k \equiv \xi (B^2_{\rm PWN} / 8 \pi)$.
Setting the injection scale of the turbulence, $k_{\rm min} \approx R^{-1}_{\rm PWN}$, we have $\nu_{{\rm res},k} = \xi (\pi/6) (c^3/r_{\rm g} R_{\rm PWN}^2)^{1/3} \propto \gamma^{-1/3}$.
The diffusion break energy is obtained by equating the diffusion length, $\sqrt{6 D_i(\gamma) t_{\rm age}}$, with the size of the nebula, $R_{\rm PWN}$, where $i = {\rm Bohm}, \parallel, \perp$.
The corresponding synchrotron photon break energies are
\begin{eqnarray}
    \varepsilon_{\rm Bohm}
    & = &
    0.00432~{\rm keV}\cdot \zeta^{-2} \left(\frac{B_{\rm PWN}}{10 ~\mu G}\right)^3 \left(\frac{R_{\rm PWN}}{{\rm pc}}\right)^4 \left(\frac{t_{\rm age}}{{\rm kyr}}\right)^{-2}, \\
    \varepsilon_{\parallel}
    & = &
    6.20 \times 10^{-15}~{\rm keV}\cdot \xi^{6} \left(\frac{B_{\rm PWN}}{10 ~\mu G}\right)^3 \left(\frac{R_{\rm PWN}}{{\rm pc}}\right)^8 \left(\frac{t_{\rm age}}{{\rm kyr}}\right)^{-6}, \\
    \varepsilon_{\perp}
    & = &
    1.60 \times 10^{1}~{\rm keV}\cdot \xi^{-6/5} \left(\frac{B_{\rm PWN}}{10 ~\mu G}\right)^3 \left(\frac{R_{\rm PWN}}{{\rm pc}}\right)^{16/5} \left(\frac{t_{\rm age}}{{\rm kyr}}\right)^{-6/5}.
\end{eqnarray}
The same equation, $\sqrt{6 D_i(\gamma) t_{\rm esc}} = R_{\rm PWN}$, gives the timescale of particle escape, and then we can find the amount of the spectral break in the same manner as Equation (\ref{eq:SynBreakSize}) by setting $t_{\rm esc} \ll t_{\rm age}, t_{\rm syn}$ in Equation (\ref{approx}), that is, $N(\gamma) \approx Q_{\rm inj} t_{\rm esc}$ in this case.
We have
\begin{eqnarray}
    \Delta \Gamma_{\rm Bohm}  & = & 0.5, \\
    \Delta \Gamma_{\parallel} & = & 0.17, \\
    \Delta \Gamma_{\perp}     & = & 0.83.
\end{eqnarray}
$\varepsilon_{\rm syn},~\varepsilon_{\rm Bohm}$, and $\varepsilon_{\perp}$ for each object are summarized in Table \ref{tbl:TypicalPhotonEnergy}, where $B_{\rm PWN}$, $t_{\rm age}$ and $R_{\rm PWN}$ are also adopted from \citet{zhu2018}.
The values of $\varepsilon_{\parallel}$ are not shown in Table \ref{tbl:TypicalPhotonEnergy} because they are too small in the radio band below megahertz frequencies.
None of the values of $\varepsilon_{\parallel}$ fits to $E_{\rm br}$ with the same values of $\zeta$ and $\xi$ for each object in our sample.

The estimates of typical photon energies in Table \ref{tbl:TypicalPhotonEnergy} are based on a one-zone model, although one-dimensional models do not dramatically change this picture \citep{kennel1984b, reynolds2009, ishizaki2017, ishizaki2018}.
The particle escape by perpendicular diffusion is the most probable candidate among these models.
The above estimate of $\varepsilon_{\perp}$ still has significant uncertainties; for example, the value of $\xi$ and the approximation $\sqrt{6 D_{\perp}(\gamma) t} = R_{\rm PWN}$, where the right-hand side is smaller than $R_{\rm PWN}$.
Further observations, especially of hard X-rays, will give us the shape of the cutoff more precisely for PWNe with various $\dot{E}$ and $\tau_c$ values, and thus will show a precise $\Delta\Gamma$ distribution and resolve which scenario is more plausible.

\begin{table}
  \tbl{
  Typical Cutoff and Break Photon Energies.\footnotemark[$*$]
  }{%
  \begin{tabular}{lcccc}
      \hline
      Symbols & N157B & PSR~J1813$-$1749 & PSR~J1400$-$6325 & G21.5$-$0.9 \\
      \hline
      \multicolumn{5}{c}{Spectral cutoff}\\
      \hline
      $E_{\rm cut}$\footnotemark[$**$]                (keV) & 67.3 ($>$32.6)              & 22.12 (16.9--31.2)             & 46.0 (30.0-94.1)              & 34.3 (32.4--36.4) \\
      $\varepsilon_{\rm size}$ (keV) & $1.58 \times 10^5$ & 684                & 305                & $3.53 \times 10^4$ \\
%       $\varepsilon_{\rm PC}$   (keV) & $1.10 \times 10^6$ & $3.92 \times 10^4$ & $5.80 \times 10^4$ & $2.47 \times 10^5$ \\
      \hline
      \multicolumn{5}{c}{Spectral break}\\
      \hline
      $E_{\rm br}$\footnotemark[$**$]  (keV)                                      & 7.4 (5.4--11.4)                  & 14.0 (8.1-- 16.6)                & 4.2 (3.9--5.2)                  & 7.1 (6.8--7.4)                  \\
      $\varepsilon_{\rm syn}$  (keV)                      & $7.68 \times 10^{-3}$ & $9.07 \times 10^{-1}$ & $5.84 \times 10^{-1}$ & $6.21 \times 10^{-3}$ \\
      $\varepsilon_{\rm Bohm}$\footnotemark[$\dagger$] (keV)    & $1.02 \times 10^{-2}$ & $6.68 \times 10^{-1}$ & $1.34 \times 10^{-1}$ & 17.5                  \\
%      $\varepsilon_{\parallel}$(keV)&                    &                       &                       &                       &                       \\
      $\varepsilon_{\perp}$\footnotemark[$\ddagger$] (keV) & 16.7                  & 182                   & 55.0                  & $6.73 \times 10^{3}$  \\
       $\Delta\Gamma_{\rm syn}$                           & 0.15 (0.04--0.26)                  & 0.71 (0.23--1.19)   & 0.34 (0.21--0.47)                 & 0.40 (0.38--0.42)                 \\
      \hline
   \end{tabular}}\label{tbl:TypicalPhotonEnergy}
\begin{tabnote}
\footnotemark[$*$] $B_{\rm PWN}, \ t_{\rm age}, \ r_{\rm TS}$, and $R_{\rm PWN}$ for each object are adopted from a broadband study of PWNe by \citet{zhu2018}.\\
\footnotemark[$**$]Same as those in Table~\ref{tab:results}.\\
\footnotemark[$\dagger$] Setting $\zeta = 1$.\\
\footnotemark[$\ddagger$] Setting $\xi = 1$.
\end{tabnote}
\end{table}

\subsection{Individual Characteristics of Pulsars and their Spectral Break}

\begin{figure}
    \centering
    \includegraphics[width=0.49\textwidth]{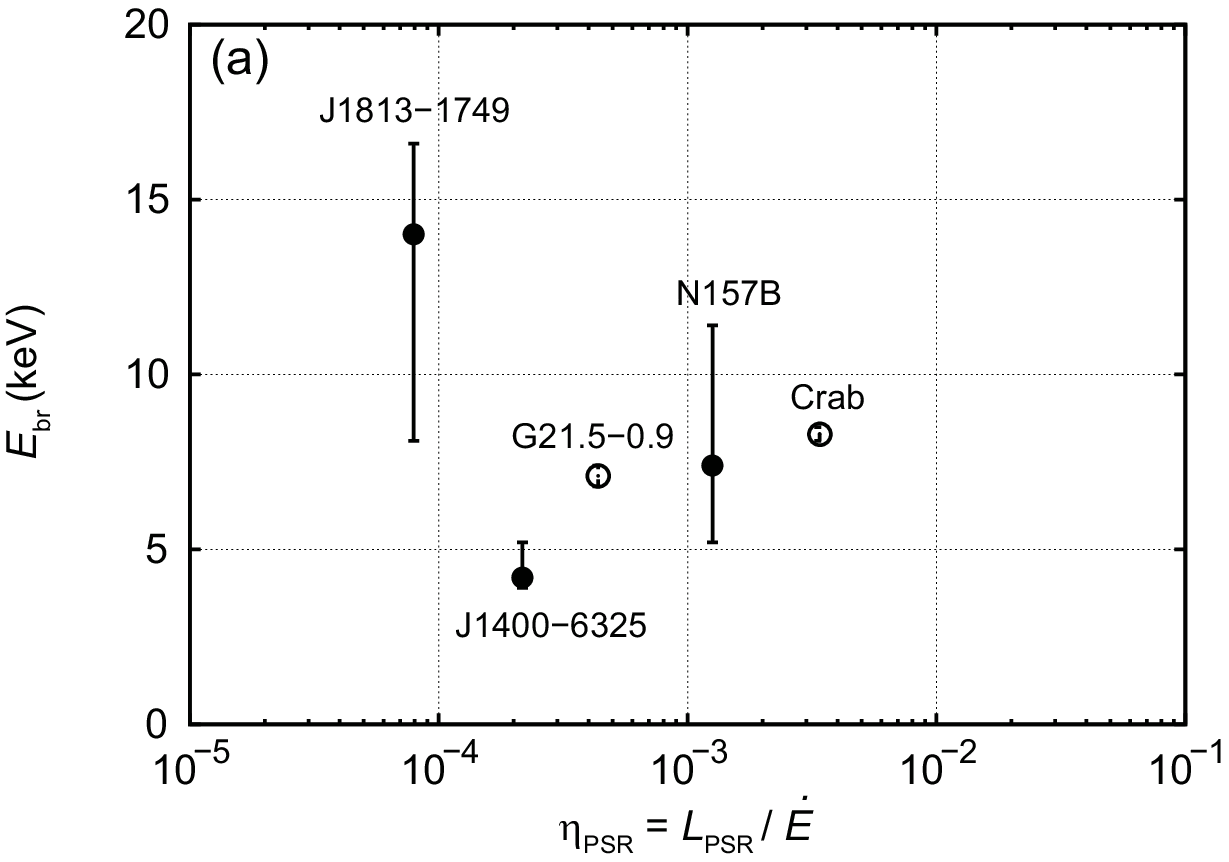}
    \includegraphics[width=0.49\textwidth]{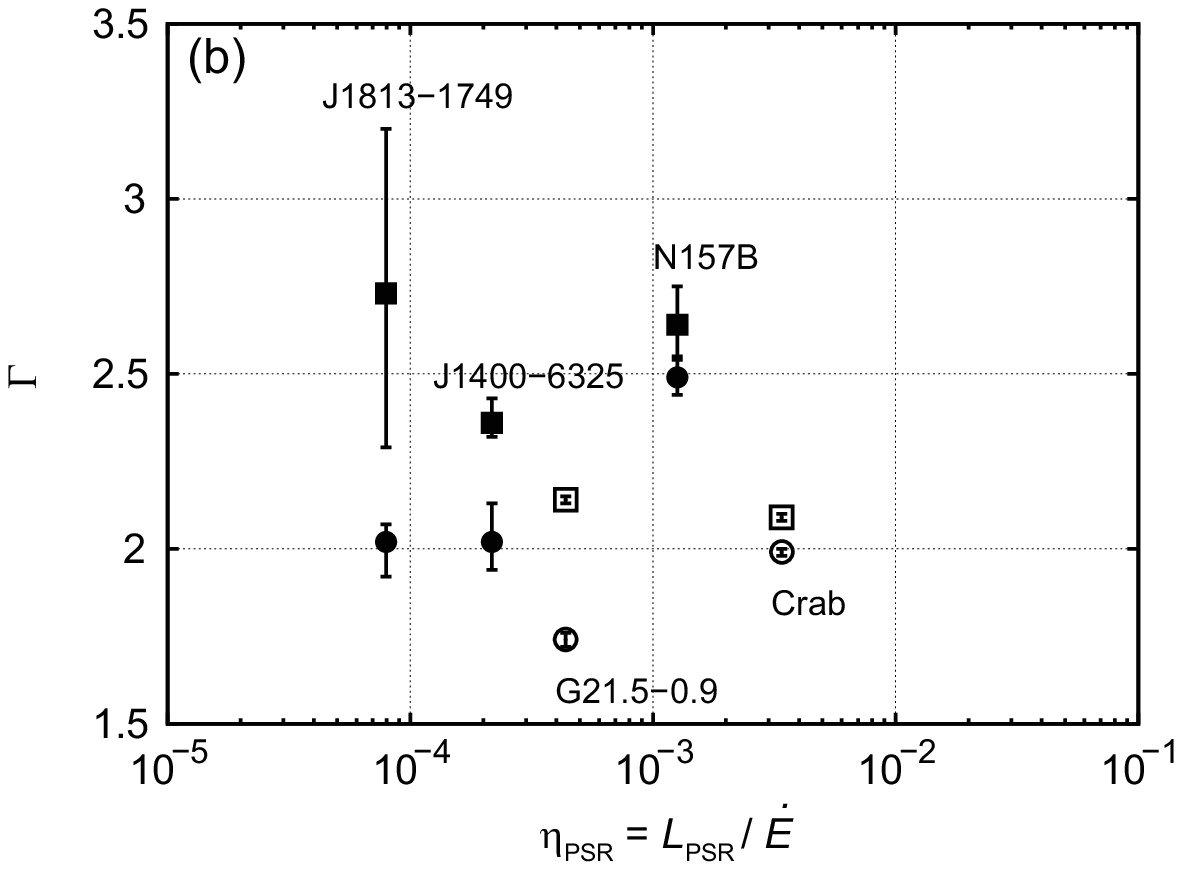}
    \includegraphics[width=0.49\textwidth]{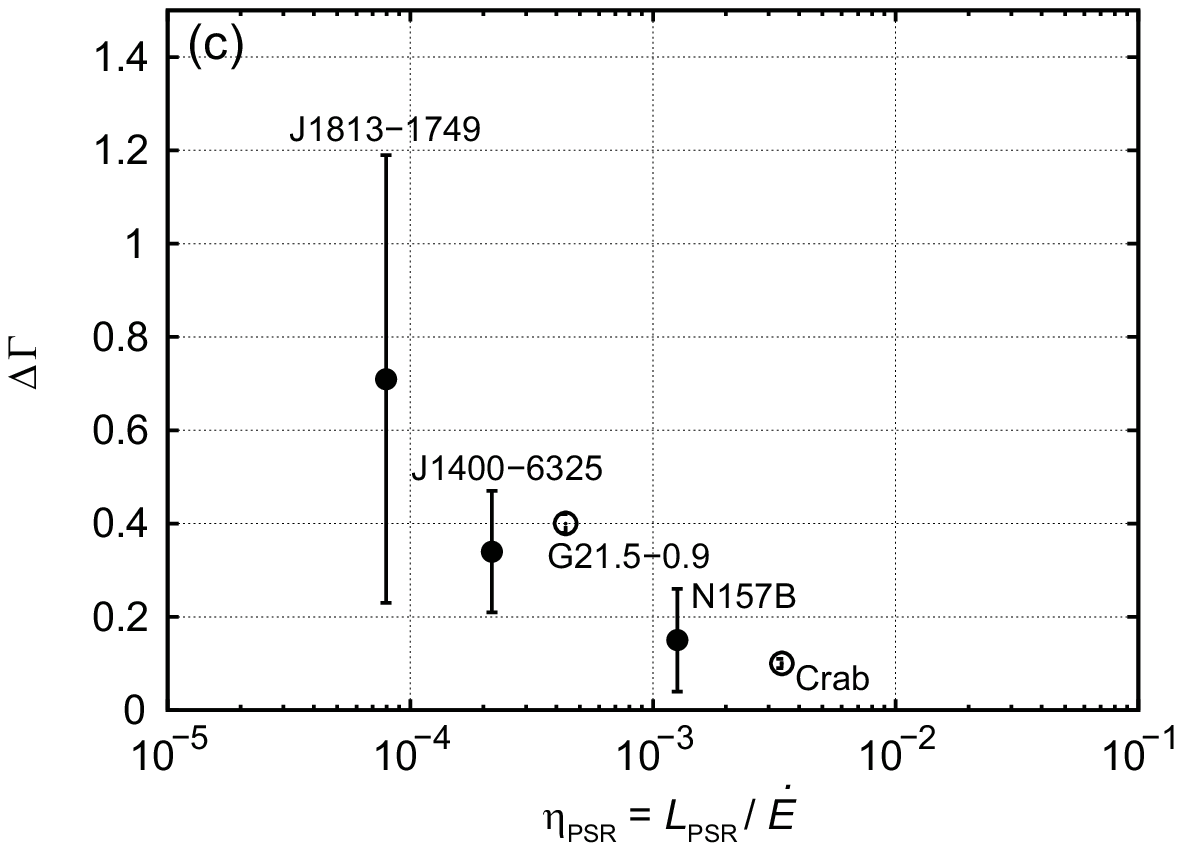}
    \includegraphics[width=0.49\textwidth]{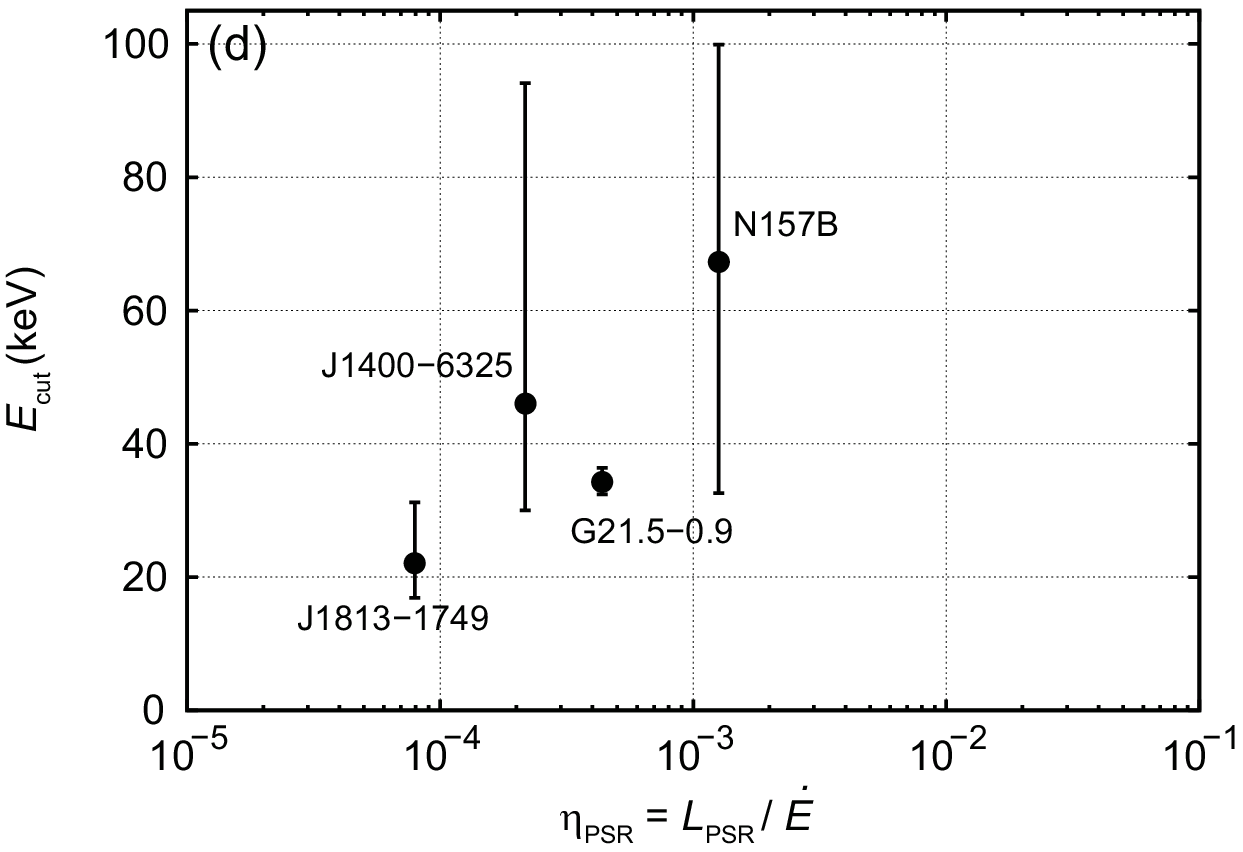}
    \caption{Relationship between $\eta_{\rm PSR} \equiv L_{\rm PSR}/\dot{E}$ and the derived parameters.
    Filled data points show our results and open data points show results from references \citep{madsen2015,hitomi2018}.
    In panel (b), circles indicate $\Gamma_1$ data and boxes indicate $\Gamma_2$ data.}
    \label{fig:psreta}
\end{figure}

In the previous section, 
we showed that 
no spectral parameters show significant correlation with $\dot{E}$ and $\tau_c$.
This result may mean that the break or cutoff feature is connected to other individual characteristics of the hosting pulsars.
One such characteristic is shown by \citet{kargaltsev2008};
the pulsar properties are also related
to the X-ray efficiency, $\eta_{\rm PSR} \equiv L_{\rm PSR}/\dot{E}$,
%which is the X-ray emitting efficiency from the spin-down energy,
where $L_{\rm PSR}$ is the X-ray luminosity of the pulsar
(see also \cite{shibata2016}).
Thus, we plot $\eta_{\rm PSR}$ against the parameters we derived.
Here, we use $L_{\rm PSR}$ values from \citet{kargaltsev2008}, which are the pulsar luminosity in the 0.5--8~keV band.
Figure~\ref{fig:psreta} shows the relation between
$\eta_{\rm PSR}$ and our parameters,
the correlation coefficients between the pulsar parameters and $\eta_{\rm PSR}$ is summarized in the third column of Table~\ref{tab:correlation}.
$\Gamma_1$ and $\Delta\Gamma$ show possible positive/negative correlations.
There is a hint of a negative correlation between $\Delta\Gamma$ and $\eta_{\rm PSR}$,
at the face value of -0.93, but this is not statistically significant and is not conclusive.
Although it is not significant, these results may imply that
the X-ray spectral break or cutoff of PWNe is
controlled by the X-ray efficiency of the central pulsars.

The X-ray luminosity of the PWN, $L_{\rm PWN}$, is positively correlated with the spin-down energy, described roughly as
$L_{\rm PWN} \propto \dot{E}$, 
but its significance is weak due to the large scatter
\citep{kargaltsev2008,li2008,kargaltsev2012,shibata2016}.
The efficiency, $\eta_{\rm PWN} = L_{\rm PWN} / \dot{E}$, ranges from $10^{-5}$ to $10^{-1}$.
The scatter is too large to attribute to incorrectly determined distance.
The X-ray luminosity of the central pulsar magnetosphere, $L_{\rm PSR}$, also shares 
the same property that it weakly correlates with $\dot{E}$ and has a large scatter about
the general trend, $L_{\rm PSR} \sim 10^{-3} \dot{E}$.
An intriguing finding is that
$\eta_{\rm PWN}$ and $\eta_{\rm PSR}$ are positively correlated, that is,
their scatter is not random but synchronized
\citep{kargaltsev2008,li2008,shibata2016}.
This implies
that the efficiency variations in the PWNe and in the magnetosphere originate from
a common underlying physical mechanism.
Another observation consistent with this argument is the correlation between the photon index of PWNe
and $\eta_{\rm PWN}$ \citep{kargaltsev2008,li2008}.
The X-ray efficiency might depend on the inclination between the magnetic
dipole axis and the rotation axis, but previous statistical studies
\citep{kargaltsev2008,li2008,vink2011,shibata2016}
do not claim a correlation between the X-ray efficiency and the inclination.
\citet{kargaltsev2012} point out that
the orthogonal rotator B0906-49 \citep{kramer2008} has an unremarkable
X-ray efficiency compared with other pulsars with similar $\dot{E}$.
A possible factor causing the individuality of the pulsars is a higher magnetic moment,
namely, the local magnetic fields near the surface. 
The strength of the local magnetic fields and their geometry 
would strongly affect
the particle acceleration and pair multiplicity.
For a given $\dot{E}$, a larger multiplicity results in a small Lorentz factor for
a wind particle injected into the nebula, and thus would affect the nebula properties.

These points lead us to examine the correlations between the spectral properties of the PWNe 
and $\eta_{\rm PSR}$. 
Moreover, because our sample lies in a small range of $\dot{E}$, it is
more likely to find a correlation with $\eta_{\rm PSR}$ 
rather than $\dot{E}$ if the correlation exists.
There is a suggestion of a correlation between $\Delta\Gamma$ and
$\eta_{\rm PSR}$ in Figure~\ref{fig:psreta}.
We do not know 
what the correlation between $\Delta\Gamma$ and $\eta_{\rm PSR}$ indicates.
%and between $E_{\rm cut}$ and $\eta_{\rm PSR}$ as well.
Further studies are required to
investigate the common mechanism that determines the efficiencies of the PWNe 
and the magnetosphere.

Our results are derived from a small number of samples.
N157B does not show a significant break or cutoff,
which may be detected significantly with better statistics and observations over a wider energy range.
Our results imply that
low X-ray efficiency pulsars may have a
clearer break in their spectrum,
and thus fainter hard X-rays.
To confirm our scenario,
we need more samples observed with better sensitivity above 10~keV.
High sensitivity in the hard X-ray band also helps us to distinguish
whether the spectral cutoff of PWNe is like a broken power-law or like a
cutoff power-law.
We need missions with high sensitivity in the hard X-ray band,
such as the FORCE mission \citep{mori2016}.

\section{Conclusion}

We conducted wideband X-ray spectroscopy of four energetic PWNe with Suzaku, Chandra, NuSTAR, and Hitomi data,
and we found that all the targets except for N157B had spectra with a significant break or cutoff
in the hard X-ray band.
We compared these derived spectral parameters with the pulsar parameters, such as $\dot{E}$ and $\tau_c$,
and no spectral parameters showed significant correlation with either of them.
A possible correlation was found between $\Delta\Gamma$, the difference in the photon index before and after the break in the broken power-law model, and the X-ray efficiency, $\eta_{\rm PSR}$, although the significance was not high enough to be conclusive. 
Although it is not significant,
the possible correlation may imply that
the break in the X-ray spectra of PWNe can be controlled by the X-ray efficiency of hosting pulsars.

\begin{ack}
We thank the anonymous referee for their fruitful comments.
We also thank Keith Arnaud and Eric Miller for their comments on the statistical aspect of our work.
This work is supported in part by Grants-in-Aid for Scientific Research
from the Japanese Ministry of Education, Culture, Sports, Science and
Technology (MEXT) of Japan, Nos. 18H05459 (AB), 19K03908 (AB), 18H01246 (SS), 21H01095 (KM), JP20K04009 (YT), and 21J01450 (WI).
S.J.T. thanks the Aoyama Gakuin University-Supported
Program "Early Eagle Program", the Sumitomo Foundation, and the Research Foundation
For Opto-Science and Technology for support.
\end{ack}

]
%\appendix 
%\section*{Case of single paragraph}

%%%
% See the manual for the detail.
%%%

\end{document}